\documentclass[11pt,a4paper]{article} 
\pdfoutput=1
\usepackage{jheppub,hyperref,mdwlist}
\usepackage{amsmath}
\usepackage{graphicx}
\usepackage{subfigure}
\usepackage{amssymb}
\usepackage{xcolor}
\usepackage{multirow}
\usepackage{cancel}
\usepackage{color}
\usepackage{ulem}
\usepackage{listings}
\usepackage{xcolor}
\usepackage{float}
\usepackage{slashed}

\usepackage{amsmath}
\usepackage{wrapfig}

\def\equ{equation}

\newcommand{\be}{\begin{\equ}}
\newcommand{\ee}{\end{\equ}}

\newcommand{\tr}{\text{tr}}
\newcommand{\Br}{\text{Br}}

\newcommand{\tabincell}[2]{\begin{tabular}{@{}#1@{}}#2\end{tabular}}
\newcommand{\Eq}[1]{Eq.~(\ref{#1})}

\newcommand{\Dfbd}{\mathord{\buildrel{\lower3pt\hbox{$\scriptscriptstyle\leftrightarrow$}}\over {D}_{\mu}}}
\newcommand{\ave}[1]{\left\langle #1\right\rangle}
\newcommand{\abs}[1]{\left| #1\right|}

\def\hc{{\rm h.c.}}

\def\mG{\mathcal{G}}
\def\mH{\mathcal{H}}

\def\mL{\mathcal{L}}

\def\mO{\mathcal{O}}

\def\mQ{\mathcal{Q}}

\def\mT{\mathcal{T}}

\def\Z{\mathbb{Z}}

\def\0{\textbf{0}}
\def\1{\textbf{1}}
\def\2{\textbf{2}}
\def\3{\textbf{3}}
\def\4{\textbf{4}}
\def\5{\textbf{5}}
\def\6{\textbf{6}}
\def\7{\textbf{7}}
\def\8{\textbf{8}}
\def\9{\textbf{9}}

\begin{document}

\title{Electroweak phase transition with composite Higgs models: calculability, gravitational waves and collider searches}

\author[1]{Ligong Bian,}
\author[2]{Yongcheng Wu,}
\author[3]{Ke-Pan Xie }

\affiliation[1]{Department of Physics, Chongqing University, Chongqing 401331, China}
\affiliation[2]{Ottawa-Carleton Institute for Physics, Carleton University, 1125 Colonel By Drive, Ottawa, Ontario K1S 5B6, Canada}
\affiliation[3]{Center for Theoretical Physics, Department of Physics and Astronomy, Seoul National University, Seoul 08826, Korea}

\emailAdd{lgbycl@cqu.edu.cn}
\emailAdd{ycwu@physics.carleton.ca}
\emailAdd{kpxie@snu.ac.kr}

\abstract{
We study the strong first order electroweak phase transition (SFOEWPT) with the $SO(6)/SO(5)$ composite Higgs model, whose scalar sector contains one Higgs doublet and one real singlet. Six benchmark models are built with fermion embeddings in \textbf{1}, \textbf{6}, and \textbf{15} of $SO(6)$. We show that SFOEWPT cannot be triggered under the {\it minimal Higgs potential hypothesis}, which assumes the scalar potential is dominated by the form factors from the lightest composite resonances. To get a SFOEWPT, the contributions from local operators induced by physics above the cutoff scale are needed. We take the $\textbf{6}+\textbf{6}$ model as an example to investigate the gravitational waves prediction and the related collider phenomenology.
}

\maketitle
\flushbottom

\section{Introduction}

The composite Higgs models (CHMs) were originally proposed to solve the Standard Model (SM) hierarchy problem. In CHMs, the Higgs boson is a composite object emerged as a pseudo-Nambu-Goldstone boson (pNGB) from the global symmetry breaking $\mG/\mH$ of a new strongly interacting sector. The interactions between the elementary (SM) sector and the composite (strong) sector break $\mG$ explicitly and generate the Higgs potential at loop levels~\cite{Kaplan:1991dc,Contino:2003ve,Agashe:2004rs}, triggering the electroweak symmetry breaking (EWSB). The Higgs boson is then naturally light. In addition, the linear mixing between the elementary quarks and the strong fermionic operator (the so-called partial compositeness mechanism) provides an explanation for the quark mass hierarchy~\cite{Agashe:2004rs}. Depends on different choices of $\mG/\mH$ and various embeddings of the SM fermions, one can have different kinds of CHMs. For example, the minimal CHM (MCHM) is based on $SO(5)/SO(4)$~\cite{Agashe:2004rs,Contino:2006qr}, which gives exactly one Higgs doublet; while the next-to-minimal CHM (NMCHM) is based on $SO(6)/SO(5)$~\cite{Gripaios:2009pe}, whose scalar sector contains one Higgs doublet and one real singlet~\footnote{In the concept of global symmetry breaking pattern, NMCHM is the minimal {\it extension} of MCHM. However, concerning about the underlying theory of the strong sector, NMCHM is the {\it minimal} model with a fundamental UV description from the bound states of new fermions. This is because $SO(6)/SO(5)\cong SU(4)/Sp(4)$, a coset that can be realized by a QCD-like theory with four-flavor Weyl fermions~\cite{Cacciapaglia:2014uja,Barnard:2013zea,Ferretti:2013kya,Hietanen:2014xca}.}.

During the last decade, people were aware that CHMs can also account for the astrophysics phenomena beyond the scope of SM. For example, Refs.~\cite{Frigerio:2012uc,Marzocca:2014msa,Ma:2017vzm,Cacciapaglia:2018avr,Cai:2018tet} consider the extra pNGBs in the non-minimal CHMs as dark matter candidates, while Refs.~\cite{Bruggisser:2018mus,Bruggisser:2018mrt,Espinosa:2011eu,Chala:2016ykx,Chala:2018opy} use CHMs to explain the baryon asymmetry of the universe. 
In the latter case, the extra scalars, either from the dilaton of the conformal invariance breaking~\cite{Bruggisser:2018mus,Bruggisser:2018mrt} or from the pNGBs of the $\mG/\mH$ global symmetry breaking~\cite{Espinosa:2011eu,Chala:2016ykx,Chala:2018opy} of the strong sector, assist the Higgs field to trigger a strong first order electroweak phase transition (SFOEWPT), creating the departure from thermal equilibrium in the early universe; while the Yukawa interactions in the quark sector provide necessary CP violating phase to realize the EW baryogenesis mechanism~\cite{Bruggisser:2018mus,Bruggisser:2018mrt,Espinosa:2011eu,Chala:2016ykx,Chala:2018opy}.

In this article, we focus on the SFOEWPT scenario of NMCHM. SFOEWPT is not only a necessary ingredient of the EW baryogenesis mechanism but also testable via gravitational waves signals at the future detectors such as LISA~\cite{Audley:2017drz}, Tianqin~\cite{Luo:2015ght}, Taiji~\cite{Hu:2017mde}, BBO~\cite{Crowder:2005nr} or DECIGO (Ultimate DECIGO)~\cite{Kawamura:2011zz,Kawamura:2006up}~\footnote{For a recent review of the cosmic phase transition and gravitational waves, see Ref.~\cite{Mazumdar:2018dfl}.}. The scalar sector of NMCHM is similar to the real singlet extensions of SM (with a $\Z_2$ symmetry in the scalar potential), which are motivated by EW baryogenesis and dark matter~\cite{Cline:2012hg,Alanne:2014bra,Huang:2015bta,Vaskonen:2016yiu,Huang:2017kzu,Huang:2018aja,Chiang:2018gsn,Bian:2018bxr,Bian:2018mkl,Cheng:2018axr,Kurup:2017dzf}. However, NMCHM differs from those models in several important aspects. First, due to the pNGB nature, the interactions between the singlet and the Higgs boson include derivative vertexes. Second, the scalar potential is not added by hand but generated by the $SO(6)$-breaking terms. Third, as a strongly interacting theory, NMCHM contains additional vector and fermion resonances, whose masses are expected to be $\mO$(TeV). Compare to previous studies about the SFOEWPT in non-minimal CHMs~\cite{Espinosa:2011eu,Chala:2016ykx}, the novelty of our work is that we consider various fermion embeddings, perform the concrete calculation of the form factor contributions to the scalar potential and point out that they are {\it not} sufficient for a SFOEWPT (see below). 

This paper is organized as follows. In Section~\ref{sec:model}, we briefly introduce the NMCHM, list the form of its scalar potential at both zero and finite temperature, and give the conditions for SFOEWPT. A complete analysis of the scalar potential is given in Section~\ref{sec:embeddings}, where we specify two sources of the scalar potential~\footnote{In principle, the scalar potential of a composite Higgs model with UV completion can be evaluated via lattice calculation. However, due to the complexity of the calculation, only very few lattice results are available for specific UV models. While no dedicated lattice calculations for the scalar potential have been done for the models mentioned in our paper, we therefore use the bottom-up approach and the form factor integrals to derive the scalar potential. This is inspired by the successful experiences in QCD (such as the calculation of the pion mass difference~\cite{Contino:2010rs,Shifman:1978bx,Shifman:1978by,Knecht:1997ts}).}: the IR contributions, which are from the one-loop form factors of the lightest composite resonances and {\it calculable}; the UV contributions, which are from local higher dimensional operators and {\it incalculable}. In many previous studies, the authors assume the UV contributions are negligible due to some unknown mechanisms of the underlying theory~\cite{Marzocca:2012zn,Pomarol:2012qf,Redi:2012ha,Marzocca:2014msa,Banerjee:2017qod}. This is known as the {\it minimal Higgs potential hypothesis} (MHP), first clearly proposed in~Ref.~\cite{Marzocca:2012zn}. However, in this study we will show that MHP is {\it not} sufficient for the SFOEWPT in NMCHM, at least for various fermion embeddings from \1 up to \1\5 representations of $SO(6)$. To trigger a SFOEWPT, we have to add the UV contributions, whose sizes are estimated by the na\"ive dimensional analysis (NDA)~\cite{Panico:2011pw}. Section~\ref{sec:6+6SFOEWPT} demonstrates that when combining the IR and UV contributions, SFOEWPT in the $\6+\6$ NMCHM can be triggered and experimentally tested by the gravitational waves. A brief discussion about the collider phenomenology of the model is also provided. Finally, we summarize and conclude in Section~\ref{sec:conclusion}.

\section{NMCHM and the SFOEWPT condition}\label{sec:model}

\subsection{A brief introduction to NMCHM}

Since we are interested in the physics at $\mO$(TeV), which is well below the confinement scale of the strong sector, the relevant physical degrees of freedom are the pNGBs, the vector and fermion resonances. In this case, the Coleman-Callan-Wess-Zumino (CCWZ) formalism~\cite{Coleman:1969sm,Callan:1969sn} can be used to describe the effective Lagrangian of NMCHM~\footnote{An excellent introduction of CCWZ application to CHM can be found in Ref.~\cite{Panico:2015jxa}.}. The full expressions and formulae are put in Appendix~\ref{app:CCWZ}, while here we only quote the main results.

Denote the 15 generators of $SO(6)$ as $T^A=\{T^{\bar A},\hat T_2^r\}$, where $T^{\bar A}=\{T_L^a,T_R^a,\hat T_1^i\}$ are the 10 generators of the unbroken $SO(5)$ [in which $\{T_L^a,T_R^a\}$ belong to the subgroup $SO(4)\cong SU(2)_L\times SU(2)_R$ while $\hat T_1^i$ belong to the coset $SO(5)/SO(4)$], and $\hat T_2^r$ are the 5 broken generators of $SO(6)/SO(5)$. The ranges of the subscripts are ($a=1,2,3$), ($i=1,...,4$) and ($r=1,...,5$). The Goldstone matrix is defined as
\be
U(\vec{\pi})=e^{i\frac{\sqrt2}{f}\pi_r\hat T^r_2},
\ee
where $f$ is the decay constant, and $\vec{\pi}=(\pi_1,...,\pi_5)^T$ are the 5 pNGBs, which transform as the \5 representation of the unbroken $SO(5)$. Under the group decomposition of $SO(5)\to SO(4)\cong SU(2)_L\times SU(2)_R$, $\5\to\4\oplus\1\cong(\2,\2)\oplus(\1,\1)$, where
\be
H=\frac{1}{\sqrt{2}}\begin{pmatrix}\pi_2+i\pi_1\\ \pi_4-i\pi_3\end{pmatrix},
\ee
is the Higgs doublet $(\2,\2)$ and $\pi_5$ is the real singlet $(\1,\1)$. Choosing the $SO(5)$-preserved vacuum state vector as $\Sigma_0=(0,0,0,0,0,1)^T$, we define the Goldstone vector as $\Sigma(\vec\pi)=U(\vec\pi)\Sigma_0$. The $d$ and $e$ symbols are given by the Maurer-Cartan form
\be
U^\dagger iD_\mu U=d_\mu^r \hat T_2^r+e^{\bar A}_\mu T^{\bar A}\equiv d_\mu+e_\mu,
\ee
where
\be\label{A_definition}
D_\mu=\partial_\mu-ig_0A_\mu=\partial_\mu-ig_0W_\mu^aT_L^a-ig'_0 B_\mu T_R^3,
\ee
is the gauge covariant derivative. We only gauge a subgroup $SU(2)_L\times U(1)_Y\subset SO(6)$, with $Y=T_R^3$.

It is convenient to work under the unitary gauge, where $\pi_{1,2,3}=0$ and $\pi_{4,5}$ are redefined as~\cite{Gripaios:2009pe}
\be\label{h_redefinition}
\frac{h}{f}=\frac{\pi_4}{\sqrt{\pi_4^2+\pi_5^2}}\sin\frac{\sqrt{\pi_4^2+\pi_5^2}}{f},\quad\frac{\eta}{f}=\frac{\pi_5}{\sqrt{\pi_4^2+\pi_5^2}}\sin\frac{\sqrt{\pi_4^2+\pi_5^2}}{f}.
\ee
Under the unitary gauge, the kinetic term of the Goldstone fields is 
\begin{multline}\label{NGB_kinetic}
\mL_{\rm kin}=\frac{f^2}{4}\tr\left[d_\mu d^{\mu}\right]=\frac{1}{2}\partial_\mu h\partial^\mu h+\frac{1}{2}\partial_\mu\eta\partial^\mu\eta \\+\frac12\frac{(h\partial_\mu h+\eta\partial_\mu\eta)^2}{f^2-h^2-\eta^2}+\frac{g_0^2}{8}h^2\left[\left(W_\mu^1\right)^2+\left(W_\mu^2\right)^2+\left(W_\mu^3-\frac{g'_0}{g_0}B_\mu\right)^2\right],
\end{multline}
in which we can read the $W$ and $Z$ mass terms after EWSB, i.e. $\ave{h}=v$. Higher order operators can also be constructed using the $d$ and $e$ symbols.

There are two kinds of composite resonances in the NMCHM. One is spin-1, similar to the $\rho$-mesons in the QCD; the other is spin-1/2, also known as the top partner. The composite objects transform in the representations of the unbroken $SO(5)$. For the vector resonances, we consider the \1\0 and \5 representations, and denote them as $\rho_\mu=\rho_\mu^{\bar A}T^{\bar A}$ and $a_\mu=a_\mu^r\hat T_2^r$. Under the decomposition $SO(5)\to SU(2)_L\times U(1)_Y$, the $\rho_\mu$ decomposes to 1 triplet, 3 singlet and 1 complex doublet; while the $a_\mu$ decomposes to 1 complex doublet and 1 singlet, i.e.
\be\label{vector_decomposition}
\begin{bmatrix}\1\0\to\3_0\oplus\1_1\oplus\1_0\oplus\1_{-1}\oplus\2_{1/2}\oplus \2_{-1/2}\\ \rho^{\bar A}\to\rho_L\oplus\rho_R^+\oplus\rho_R^0\oplus\rho_R^{-1}\oplus\rho_D\oplus\tilde\rho_D\end{bmatrix};\quad
\begin{bmatrix}
\5\to\2_{1/2}\oplus \2_{-1/2}\oplus\1_0\\ a^{r}\to a_D\oplus\tilde a_D\oplus a_S
\end{bmatrix},
\ee
where $\tilde\rho_D=i\sigma^2\rho_D^*$ and $\tilde a_D=i\sigma^2 a_D^*$. The full expressions of the resonances can be found in Appendix~\ref{app:CCWZ}. The Lagrangian of vector resonances reads~\footnote{In the Lagrangians of this subsection, the summation of resonances with the same quantum number is always implied, e.g.
\begin{equation*}
-\frac{1}{4}\tr\left[\rho_{\mu\nu}\rho^{\mu\nu}\right]\to-\frac{1}{4}\sum_{n=1}^{N_\rho}\tr\left[\rho_{(n)\mu\nu}\rho_{(n)}^{\mu\nu}\right],\quad
\end{equation*}
and similar for $a_\mu$, $\Psi_{\1\0}$, $\Psi_\5$ and $\Psi_\1$. Generally we assume the resonance labeled by a larger number is heavier, such as $M_{\rho(n+1)}>M_{\rho(n)}$.}
\be\label{spin-1}
\mL_\rho=-\frac{1}{4}\tr\left[\rho_{\mu\nu}\rho^{\mu\nu}\right]+\frac{M_{\rho}^2}{2g_{\rho}^2}\tr\left[(g_{\rho}\rho_\mu-e_\mu)^2\right]-\frac{1}{4}\tr[a_{\mu\nu}a^{\mu\nu}]+\frac{M_a^2}{2}\tr\left[a_\mu a^\mu\right],
\ee
where $g_\rho\gg g_0$, $g'_0$ is the coupling constant of the strong sector. The field strengths are
\be
\rho_{\mu\nu}=\partial_\mu\rho_\nu-\partial_\nu\rho_\mu-ig_\rho [\rho_\mu,\rho_\nu],\quad a_{\mu\nu}=\nabla_\mu a_\nu-\nabla_\nu a_\mu,
\ee
where the $SO(6)/SO(5)$ covariant derivative is $\nabla_\mu=\partial_\mu-ie_\mu$.

For the fermion resonances, we consider the \1, \5 and \1\0 representations of $SO(5)$. To give the correct hypercharge, an extra $U(1)_X$ is introduced, and the gauging of hypercharge is extended to $Y=T_R^3+X$. As we will see, for the fermion resonances relevant to the top-quark interactions, $X=2/3$. Under the decomposition $SO(5)\times U(1)_X\to SU(2)_L\times U(1)_Y$, we get
\be\label{fermion_decomposition}
\begin{bmatrix}
\5_{2/3}\to\2_{7/6}\oplus \2_{1/6}\oplus\1_{2/3}\\ \Psi_\5\to Q_X\oplus Q\oplus\widetilde T\end{bmatrix};\quad
\begin{bmatrix}\1\0_{2/3}\to\3_{2/3}\oplus \1_{5/3}\oplus\1_{2/3}\oplus\1_{-1/3}\oplus\2_{7/6}\oplus \2_{1/6}\\
 \Psi_{\1\0}\to Y\oplus K_{5/3}\oplus K_{2/3}\oplus K_{-1/3}\oplus J_X\oplus J_Q
\end{bmatrix},
\ee
where the full expressions of the above fields are given in Appendix~\ref{app:CCWZ}. The Lagrangian of top partners reads
\begin{multline}\label{spin-1/2}
\mL_\Psi=\tr\left[\bar\Psi_{\1\0}\left(i\slashed{\nabla}+g_0'\frac23\slashed{B}-M_{\1\0}\right)\Psi_{\1\0}\right]\\
+\bar\Psi_{\5}\left(i\slashed{\nabla}+g_0'\frac23\slashed{B}-M_\5\right)\Psi_\5+\bar\Psi_{\1}\left(i\slashed{\partial}+g_0'\frac23\slashed{B}-M_\1\right)\Psi_\1,
\end{multline}
where the $SO(6)/SO(5)$ covariant derivatives
\be
\nabla_\mu\Psi_\5=\left(\partial_\mu-ie_\mu^{\bar A}t^{\bar A}\right)\Psi_\5,\quad \nabla_\mu\Psi_{\1\0}=\partial_\mu\Psi_{\1\0}-ie_\mu^{\bar A}t^{\bar A}\Psi_{\1\0}+i\Psi_{\1\0}e_\mu^{\bar A}t^{\bar A},
\ee
and the matrices $[t_{L,R}^a]_{rs}\equiv [T_{L,R}^a]_{rs}$, $[\hat t_1^i]_{rs}\equiv [\hat T_1^i]_{rs}$ with $(r,s=1,...,5)$.

The SM fermions gain their masses through EWSB and the mixing with the strong sector, i.e. the partial compositeness interactions. The heavier a fermion is, the more strongly it couples to the top partners. Therefore, only the interactions with top quark are sizable due to the large top mass, and hereafter we only consider $q_L=(t_L,b_L)^T$ and $t_R$. In CCWZ, the elementary fermions are embedded into the incomplete representation of $SO(6)$, and one has the degree of freedom to choose various embeddings when building the model. For $q_L$, we consider the \6 and \1\5 representations; while for $t_R$, we consider the \1, \6 and \1\5 representations. The explicit expressions of the embeddings are as follows. First, $t_R$ can be the $\1_{2/3}$ of $SO(6)\times U(1)_X$: $t_R^\1\equiv t_R$. Second, under group decomposition $SO(6)\times U(1)_X \to SU(2)_L\times U(1)_Y$ we get
\be
\6_{2/3}\to \2_{7/6}\oplus\2_{1/6}\oplus\1_{2/3}\oplus\1_{2/3},
\ee
thus the embedding of $q_L$ is unique while of $t_R$ can be the superposition of the two $\1_{2/3}$: 
\be
q_L^{\6}=\frac{1}{\sqrt{2}}\begin{pmatrix}ib_L& b_L& it_L& -t_L& 0& 0\end{pmatrix}^T,\quad
t_R^{\6}=\begin{pmatrix}0&0&0&0&t_Re^{i\phi}c_\theta&t_Rs_\theta\end{pmatrix}^T,
\ee
where $c_\theta$ and $s_\theta$ stand respectively for $\cos\theta$ and $\sin\theta$, with $\theta$ and $\phi$ being the mixing angles~\cite{Redi:2012ha}. The phase $\phi$ is unphysical~\cite{Redi:2012ha,Mrazek:2011iu}.

Finally, we consider the \1\5 representation. Under the decomposition chain $SO(6)\times U(1)_X\to SU(2)_L\times U(1)_Y$ we have 
\be\label{15_decomposition}
\1\5_{2/3}\to \3_{2/3}\oplus\1_{5/3}\oplus\1_{2/3}\oplus\1_{-1/3}
\oplus\2_{7/6}\oplus\2_{1/6}\oplus\2_{7/6}\oplus\2_{1/6}\oplus\1_{2/3}.
\ee
Since two $\2_{1/6}$ are obtained, we have two different ways to embed $q_L$ in to \1\5, namely
\be\label{qL15}
q_L^{\1\5_A}=(q_L^\6)_j\hat T^{j}_1,\quad q_L^{\1\5_B}=i(q_L^\6)_j\hat T^{j}_2,
\ee
where $(j=1,...,4)$. The $\1\5_B$ embedding has been considered in Ref.~\cite{Banerjee:2017wmg}, while the $\1\5_A$ is first proposed here. Phenomenologically, the model with $\1\5_B$ embedding is stringently constrained by the $Zb_L\bar b_L$ coupling measurement, see Appendix~\ref{app:Zbb} for the details. Hereafter we only consider $q_L^{\1\5_A}$ and denote it as $q_L^{\1\5}$. On the other hand, the right-handed top can be embedded into the superposition of the two $\1_{2/3}$ in \Eq{15_decomposition}, i.e.
\be\label{tR15}
t_R^{\1\5}=T_R^3 t_Rc_\theta+\hat T_2^5t_Re^{i\phi}s_\theta.
\ee
The special case $\theta=0$ is considered in Ref.~\cite{Banerjee:2017wmg} for a collider phenomenology study. 

Having the SM embeddings in hand, we are able to write down the partial compositeness interactions. Since $q_L$ has two different embeddings while $t_R$ has three, the combinations yield six different models, which can be labeled by (left-handed embedding)$+$(right-handed embedding). For example, the $\1\5+\6$ NMCHM means the benchmark model with $q_L$ embedded in \1\5 while $t_R$ embedded in \6. We will discuss those models one by one.

\textbf{The $\6+\1$ model}:
\be\label{6+1}
\mL_{\6+\1}\supset y_L^\5 f (\bar{q}_L^{\6})_IU_{Ir}\Psi_{\5}^r +y_L^\1 f (\bar{q}_L^{\6})_IU_{I6}\Psi_{\1}+y_R^\1 f\bar{t}_R^\1\Psi_{\1}+\hc~,
\ee
where $(I=1,...,6)$. After EWSB, the Yukawa interactions give mass to the top quark.

\textbf{The $\6+\6$ model}:
\be\label{6+6}
\mL_{\6+\6}\supset y_L^\5 f (\bar{q}_L^{\6})_IU_{Ir}\Psi_{\5}^r +y_L^\1 f (\bar{q}_L^{\6})_IU_{I6}\Psi_{\1}+y_R^\5 f (\bar{t}_R^{\6})_IU_{Ir}\Psi_{\5}^r+y_R^\1  f (\bar{t}_R^{\6})_IU_{I6}\Psi_{\1} +\hc~.
\ee

\textbf{The $\6+\1\5$ model}:
\begin{multline}\label{6+15}
\mL_{\6+\1\5}\supset y_L^\5 f (\bar{q}_L^{\6})_IU_{Ir}\Psi_{\5}^r +y_L^\1 f (\bar{q}_L^{\6})_IU_{I6}\Psi_{\1}\\
+y_R^{\1\0} f (\bar{t}_R^{\1\5})_{IJ}U_{Jr}\Psi_{\1\0}^{rs}[U^\dagger]_{sI}+y_R^\5 f \Sigma^\dagger_I(\bar{t}_R^{\1\5})_{IJ}U_{Jr}\Psi_{\5}^r+\hc~.
\end{multline}

\textbf{The $\1\5+\1$ model}:
\be\label{15+1}
\mL_{\1\5+\1}\supset y_L^{\1\0} f (\bar{q}_L^{\1\5})_{IJ}U_{Jr}\Psi_{\1\0}^{rs}[U^\dagger]_{sI}+y_L^\5 f \Sigma^\dagger_I(\bar{q}_L^{\1\5})_{IJ}U_{Jr}\Psi_{\5}^r+y_R^\1 f\bar{t}_R^\1\Psi_{\1}+\hc~,
\ee
where $(I,J=1,...,6)$. Note that the Yukawa interactions in above equation cannot give a mass to the top quark, because $q_L$ mixes with $\Psi_{\1\0}$ and $\Psi_{\5}$, while $t_R$ mixes with $\Psi_\1$. Therefore, this model is not supported by reality.

\textbf{The $\1\5+\6$ model}:
\begin{multline}\label{15+6}
\mL_{\1\5+\6}\supset y_L^{\1\0} f (\bar{q}_L^{\1\5})_{IJ}U_{Jr}\Psi_{\1\0}^{rs}[U^\dagger]_{sI}+y_L^\5 f \Sigma^\dagger_I(\bar{q}_L^{\1\5})_{IJ}U_{Jr}\Psi_{\5}^r\\
+y_R^\5 f (\bar{t}_R^{\6})_IU_{Ir}\Psi_{\5}^r+y_R^\1  f (\bar{t}_R^{\6})_IU_{I6}\Psi_{\1} +\hc~.
\end{multline}

\textbf{The $\1\5+\1\5$ model}: 
\begin{multline}\label{15+15}
\mL_{\1\5+\1\5}\supset y_L^{\1\0} f (\bar{q}_L^{\1\5})_{IJ}U_{Jr}\Psi_{\1\0}^{rs}[U^\dagger]_{sI}+y_L^\5 f \Sigma^\dagger_I(\bar{q}_L^{\1\5})_{IJ}U_{Jr}\Psi_{\5}^r\\
+y_R^{\1\0} f (\bar{t}_R^{\1\5})_{IJ}U_{Jr}\Psi_{\1\0}^{rs}[U^\dagger]_{sI}+y_R^\5 f \Sigma^\dagger_I(\bar{t}_R^{\1\5})_{IJ}U_{Jr}\Psi_{\5}^r+\hc~.
\end{multline}

In summary, we get five different NMCHMs to study (the $\1\5+\1$ model is dropped because of the issue of massless top quark).

\subsection{The scalar potential and the condition of SFOEWPT}

In the strong sector, $h$ and $\eta$ are protected by the Goldstone theorem and strictly massless. It is the $SO(6)$-breaking interactions between the elementary sector and the strong sector that generate the effective potential $V(h,\eta)$. As we will see in Section~\ref{sec:embeddings}, the potential can be written in a very good approximation as
\be\label{V_scalar}
V(h,\eta)=\frac{\mu_h^2}{2}h^2+\frac{\lambda_h}{4}h^4+\frac{\mu_\eta^2}{2}\eta^2+\frac{\lambda_\eta}{4}\eta^4+\frac{\lambda_{h\eta}}{2}h^2\eta^2.
\ee
Above potential implies a $\Z_2$ symmetry $\eta\to-\eta$, which might be broken either spontaneously by the global vacuum expectation value (VEV) of $\eta$ or explicitly by the Yukawa interactions such as $t\bar t\eta$ (depends on the choice of $\theta$ in the $t_R$ embedding). A physically acceptable potential $V(h,\eta)$ should have a VEV $\ave{h}=v$ at zero temperature and give correct masses to the observed particles such as the Higgs boson, the $W^\pm$ and $Z$ bosons, the top quark, etc.

At finite temperature, $V(h,\eta)$ receives the thermal corrections and the vacuum structure changes. For the tree-level driven first-order EWPT, the high-temperature expansion approximation of the finite temperature potential could be adopted to characterize the dynamics of the phase transition~\cite{Espinosa:2011ax}.
Keeping only the leading $T^2$ terms~\footnote{Which has been proved to be gauge-independent in Refs.~\cite{Dolan:1973qd,Braaten:1989kk}.}, the finite temperature potential is then written as:
\be\label{VT_scalar}
V_T(h,\eta)=\frac{\mu_h^2+c_hT^2}{2}h^2+\frac{\lambda_h}{4}h^4+\frac{\mu_\eta^2+c_\eta T^2}{2}\eta^2+\frac{\lambda_\eta}{4}\eta^4+\frac{\lambda_{h\eta}}{2}h^2\eta^2,
\ee
where 
\be\label{ch_ceta}
c_h=\frac{3g^2+g'^2}{16}+\frac{y_t^2}{4}+\frac{\lambda_h}{2}+\frac{\lambda_{h\eta}}{12},\quad c_\eta=\frac{\lambda_\eta}{4}+\frac{\lambda_{h\eta}}{3},
\ee
with $g^{(\prime)}$ and $y_t$ being the physical EW couplings and top Yukawa, respectively. The necessary condition for SFOEWPT is the existence of two degenerate vacuums at some critical temperature $T_c$. In the thermal potential $V_T(h,\eta)$, the proper way to realize that is the so-called ``two-step'' phase transition~\footnote{We briefly comment on the other two possible SFOEWPT mechanisms. The first one is the ``one-step'' SFOEWPT, in which a potential barrier is induced only along the $h$ direction, and the $\eta$ never gets a VEV~\cite{Kurup:2017dzf}. This scenario exists only when the thermal corrections depend linearly on $T$ are included. As those terms cause gauge-dependent $T_c$ and $v_c$~\cite{Patel:2011th}, we will not consider them here. The second one is the effective field theory (EFT) scenario, in which a heavy $\eta$ is integrated out, leaving the dimensional-6 operators that generate SFOEWPT~\cite{Zhang:1992fs,Grojean:2004xa,Gan:2017mcv,Huang:2015izx,Huang:2016odd,Cao:2017oez}. However, a portal interaction $\eta h^2$ is crucial in generating a sizable $h^6$ operator~\cite{Cao:2017oez}. While such a portal term is absent in our potential \Eq{V_scalar}, the EFT scenario is disfavored.}, in which the VEV $(\ave{h},\ave{\eta})$ changed as $(0,0)\to(0,w)\to(v,0)$ when the universe cooled down from the temperature $T\gg M_h$ to $T=0$. This also tells us the $\Z_2$ symmetry of $\eta$ is preserved by the scalar potential at zero temperature (but it might be broken by the Yukawa interactions, see the discussions in Section~\ref{sec:embeddings}).

Now we address the conditions for the two-step phase transition. The method used here is similar to those in Refs.~\cite{Bian:2018mkl,Bian:2018bxr}. At zero temperature there should be a EW breaking local minimum $(v,0)$ along the $h$ direction, which requires
\be\label{h_vacuum}
\mu_h^2<0,\quad\lambda_h>0,\quad \lambda_h\mu_\eta^2>\lambda_{h\eta}\mu_h^2,\quad \Rightarrow v=\sqrt{-\mu_h^2/\lambda_h};
\ee
and another local minimum $(0,w)$ along the $\eta$ direction, which needs
\be\label{eta_vacuum}
\mu_\eta^2<0,\quad\lambda_\eta>0,\quad \lambda_\eta\mu_h^2>\lambda_{h\eta}\mu_\eta^2,\quad\Rightarrow w=\sqrt{-\mu_\eta^2/\lambda_\eta}.
\ee
Note the third inequalities in Eqs.~(\ref{h_vacuum}) and (\ref{eta_vacuum}) come from the Hessian matrix and ensure $(v,0)$, $(0,w)$ to be local minima but not saddle points. One can infer $\lambda_{h\eta} > 0$ and $\lambda_{h\eta}^2>\lambda_h\lambda_\eta$ from those inequalities too. In addition, the EWSB minimum should be the true vacuum, i.e.
\be\label{Z2_vacuum}
V(v,0)=-\frac{\mu_h^4}{4\lambda_h}<V(0,w)=-\frac{\mu_\eta^4}{4\lambda_\eta},
\ee
thus $\mu_\eta^2\sqrt{\lambda_h}>\mu_h^2\sqrt{\lambda_\eta}$.

At the critical temperature $T_c$, there should exist two degenerate vacuums $(v_c,0)$ and $(0,w_c)$ satisfying
\be\label{Z2T_vacuum}\begin{split}
&\mu_h^2+c_hT_c^2<0,\quad\lambda_h(\mu_\eta^2+c_\eta T_c^2)>\lambda_{h\eta}(\mu_h^2+c_hT_c^2),\quad v_c=\sqrt{-(\mu_h^2+c_hT_c^2)/\lambda_h};\\
&\mu_\eta^2+c_\eta T_c^2<0,\quad\lambda_\eta(\mu_h^2+c_hT_c^2)>\lambda_{h\eta}(\mu_\eta^2+c_\eta T_c^2),\quad w_c=\sqrt{-(\mu_\eta^2+c_\eta T_c^2)/\lambda_\eta},
\end{split}\ee
and
\be
V(v_c,0)=-\frac{(\mu_h^2+c_hT_c^2)^2}{4\lambda_h}=V(0,w_c)=-\frac{(\mu_\eta^2+c_\eta T_c^2)^2}{4\lambda_\eta}.
\ee
Solving the above equation yields
\be
T_c^2=\frac{\mu_h^2\sqrt{\lambda_\eta}-\mu_\eta^2\sqrt{\lambda_h}}{c_\eta\sqrt{\lambda_h}-c_h\sqrt{\lambda_\eta}}.
\ee
Requiring $T_c\in \mathbb{R}$ yields $c_h\sqrt{\lambda_\eta}>c_\eta\sqrt{\lambda_h}$. Substituting the expression of $T_c$ into \Eq{Z2T_vacuum}, one obtain $c_\eta\mu_h^2>c_h\mu_\eta^2$. Combining all the inequalities we get, the condition of two degenerate vacuums for $V_T(h,\eta)$ is
\be\label{degenerate_vacuum}
\frac{c_\eta}{c_h}<\frac{\mu_\eta^2}{\mu_h^2}<\frac{\sqrt{\lambda_\eta}}{\sqrt{\lambda_h}}<\frac{\lambda_{h\eta}}{\lambda_h}.
\ee

Note that \Eq{degenerate_vacuum} is necessary but not sufficient for a first order EWPT. To really achieve a first order EWPT, one should calculate the bubble nucleation rate per volume in the early universe
\be
\Gamma/V\approx T^4\left(\frac{S_3}{2\pi T}\right)^{3/2}e^{-S_3(T)/T},
\ee
and confirm that the critical condition
\be\label{nucleation_condition}
\frac{S_3(T_n)}{T_n}\sim 4\ln\frac{\xi M_{\rm Pl}}{T_n}\sim140,
\ee
is satisfied at some nucleation temperature $T_n$. Here $S_3$ is the classical action of the $O(3)$ symmetric bounce solution~\cite{Linde:1981zj}, $\xi\approx0.03$ and $M_{\rm Pl}=1.22\times10^{19}$ GeV. Normally $T_n$ is slightly lower than $T_c$. Only when \Eq{nucleation_condition} is satisfied can the bubbles percolate in an expanding universe and phase transition successfully complete. In addition, to avoid the generated baryon asymmetry being washed out, the EW sphaleron process should be suppressed. That means the phase transition should be sufficiently strong~\cite{Moore:1998swa,Quiros:1999jp,Zhou:2019uzq}, satisfying
\be\label{SFOEWPT_condition}
v_n/T_n\gtrsim1,
\ee
where $v_n$ is the Higgs VEV at $T_n$.
We will deal with Eqs.~(\ref{nucleation_condition}) and (\ref{SFOEWPT_condition}) numerically in Section~\ref{sec:6+6SFOEWPT}. 

In the end of this subsection, we discuss the allowed parameter space under \Eq{degenerate_vacuum}. At zero temperature, due to the derivative interactions in the kinetic term, the field shift that canonicalizes the Higgs kinetic term should be
\be
h\to v+\sqrt{1-\frac{v^2}{f^2}}h,
\ee
which changes the zero temperature potential \Eq{V_scalar} to
\be\begin{split}
V(h,\eta)\to&~-\mu_h^2\left(1-\frac{v^2}{f^2}\right)h^2+\lambda_hv\left(1-\frac{v^2}{f^2}\right)^{3/2}h^3+\frac{\lambda_h}{4}\left(1-\frac{v^2}{f^2}\right)^2h^4\\
&~+\frac{1}{2}(\mu_\eta^2+\lambda_{h\eta}v^2)\eta^2+\frac{\lambda_\eta}{4}\eta^4+\lambda_{h\eta}v\sqrt{1-\frac{v^2}{f^2}}h\eta^2+\frac{\lambda_{h\eta}}{2}\left(1-\frac{v^2}{f^2}\right)h^2\eta^2,
\end{split}\ee
and the physical masses can be easily read as
\be\label{scalar_mass}
M_h^2=-2\mu_h^2\left(1-\frac{v^2}{f^2}\right),\quad M_\eta^2=\mu_\eta^2+\lambda_{h\eta}v^2.
\ee
Since $v^2\ll f^2$ is expected, $\mu_h^2$ is almost fixed by the observed $M_h=125.09$ GeV. And $\lambda_h$ is also fixed by $-\mu_h^2/v^2$. The mass and the Higgs coupling of the EW bosons are respectively
\be
M_W^2=\frac{g^2v^2}{4},\quad g_{hWW}=\frac{g^2v}{2}\sqrt{1-\frac{v^2}{f^2}}=g_{hWW}^{\rm SM}\sqrt{1-\frac{v^2}{f^2}},
\ee
see the Goldstone kinetic term in \Eq{NGB_kinetic}. Current EW and Higgs measurements have constrained $f\gtrsim1$ TeV.

When $M_\eta <M_h/2$, the decay channel $h\to \eta\eta$ opens and the partial width is~\cite{Frigerio:2012uc}
\be
\Gamma(h\to \eta\eta)=\frac{v^2}{32\pi M_h}\left(\frac{M_h^2}{f\sqrt{f^2-v^2}}-2\lambda_{h\eta}\sqrt{1-\frac{v^2}{f^2}}\right)^2\sqrt{1-\frac{4M_\eta^2}{M_h^2}}.
\ee
Depending on the various $\eta$ decay channels, $h\to\eta\eta$ can lead to invisible decay (for the dark matter scenario), multi-boson final state (if $\eta$ decays to a pair of EW bosons via WZW anomaly) or multi-jet final state (if $\eta$ decays to $jj$ or $gg$ via fermion loops), etc. On the other hand, the Higgs total width in SM is extremely small that $\Gamma_h=4.07$ MeV~\cite{Denner:2011mq}, thus even a small $h\eta\eta$ vertex can change the Higgs branching ratios a lot. As SFOEWPT needs a sizable $\lambda_{h\eta}$ (see \Eq{degenerate_vacuum} for details), were $h\to\eta\eta$ allowed it would dominate the Higgs decay. This would be ruled out by the existing experimental measurements~\cite{ATLAS-CONF-2018-031,Sirunyan:2018koj}, which show compatible branching ratios with the SM prediction. To avoid this conflict, we will consider only the $M_\eta>M_h/2$ region, and $h\to\eta\eta$ is then forbidden by phase space.

\begin{figure}
\centering
\includegraphics[scale=0.45]{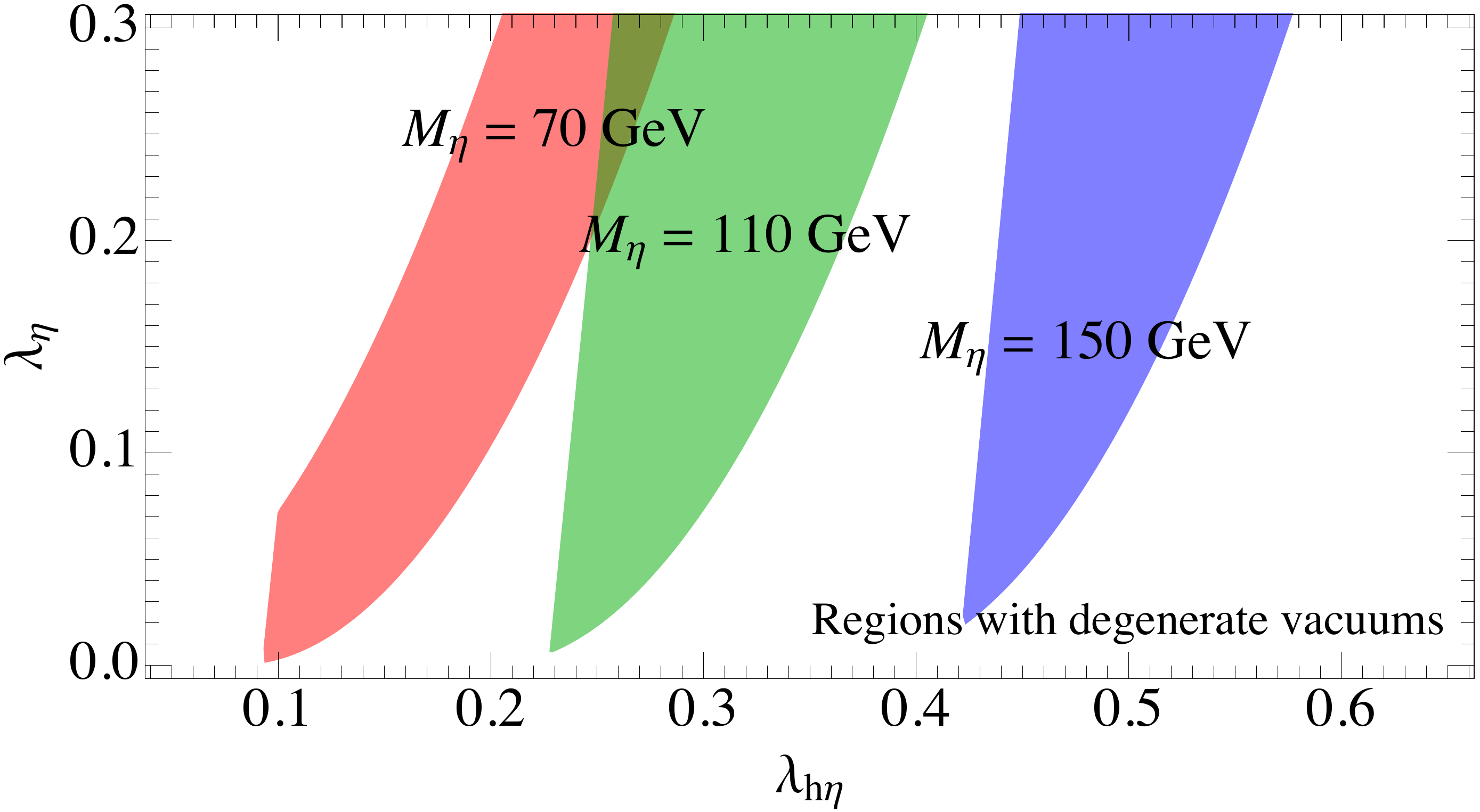}
\caption{The parameter space giving degenerate vacuums.}
\label{fig:lambda}
\end{figure}

Given \Eq{scalar_mass}, the coefficients in $V(h,\eta)$ can be expressed in terms of $f$, $M_\eta$, $\lambda_\eta$ and $\lambda_{h\eta}$, because $M_h$ and $v$ are fixed by experiments. As long as $f\gg v$, the dependence of $f$ is mild and the degrees of freedom reduce to three. In Fig.~\ref{fig:lambda} we plot the parameter regions allowed by \Eq{degenerate_vacuum} for different $M_\eta$ values. One can see that $\lambda_{h\eta}$ has a positive correlation with $M_\eta$, as expected.

\section{Deriving the scalar potential of NMCHM}\label{sec:embeddings}

In this section, we first classify the sources of the potential and then investigate them one by one. Especially, we will demonstrate that the IR contributions can't trigger SFOEWPT alone.

\subsection{The sources of the scalar potential}

The coefficients $\mu_{h,\eta}^2$ and $\lambda_{h,\eta,h\eta}$ in $V(h,\eta)$ are generated by two kinds of $SO(6)$-breaking interactions. The first type is gauge interaction, see \Eq{A_definition}. This breaks $SO(6)\times U(1)_X$ into its largest subgroup containing the SM gauge group as an ideal, i.e. $SU(2)_L\times U(1)_Y\times U(1)_\eta$~\cite{Gripaios:2009pe}, where $U(1)_\eta$ is the subgroup generated by $\hat T_2^5$. As a result, the gauge interactions contribute to the potential for $h$ (i.e. $\mu_h^2$ and $\lambda_h$) but not $\eta$ (i.e. $\mu_\eta^2$ and $\lambda_{\eta,h\eta}$).

The second source of the potential comes from the partial compositeness terms, i.e. Eqs.~(\ref{6+1}) to (\ref{15+15}). In general, they break $SO(6)\times U(1)_X$ into $SU(2)_L\times U(1)_Y$ and contribute to all coefficients in $V(h,\eta)$. However, in some embeddings, accidentally the elementary fermion multiplet has a definite $U(1)_\eta$ quantum number and then its contribution to the $\eta$ potential vanishes. For the embeddings considered in this paper, under the action of $e^{i\alpha^5\hat T_2^5}$,
\be
\delta q_L^\6=0,\quad \delta t_R^\1=0,\quad \delta t_R^{\1\5}=0.
\ee
Thus they have $U(1)_\eta$ quantum number zero~\footnote{In case $\theta=\pi/4$ and $\phi=\pi/2$, $t_R^\6$ also has a definite $U(1)_\eta$ charge $-1/\sqrt{2}$.}. As a result we expect $\mu_\eta^2$ and $\lambda_\eta$ receive no contributions from the $q_L^\6$, $t_R^\1$ and $t_R^{\1\5}$ embeddings ($\lambda_{h\eta}$ may receive contributions from the combination of one of these embeddings and another $U(1)_\eta$-breaking embedding). Hence in the $\6+\1$ and $\6+\1\5$ NMCHMs, $U(1)_\eta$ is only spontaneously broken by the strong dynamics and $\eta$ remains as an exactly massless NGB. These two models are not only unable to trigger SFOEWPT but also ruled out by the experimental searches for axion~\cite{Gripaios:2009pe,Georgi:1986df}, thus they will not be studied in the rest of this paper. In summary, only the $\6+\6$, $\1\5+\6$ and $\1\5+\1\5$ NMCHMs are considered for the SFOEWPT in the following text.

\begin{table}
\footnotesize\renewcommand\arraystretch{1.5}\centering
\begin{tabular}{|c|c|c|}
 \hline
 & Gauge-induced & Fermion-induced \\ \hline
\tabincell{c}{IR contributions\\(calculable)} & \tabincell{c}{Form factors from \Eq{spin-1},\\in terms of $g^{(\prime)},~g_\rho$ and $M_{\rho,a}$} & \tabincell{c}{Form factors from Eqs.~(\ref{spin-1/2}) and (\ref{6+1})$\sim$(\ref{15+15}),\\in terms of $M_{\1,\5,\1\0}$ and $y_{L,R}^{\1,\5,\1\0}$} \\ \hline
\tabincell{c}{UV contributions\\(estimated by NDA)} & Local operators involved $g^{(\prime)}$ & Local operators involved $y_{L,R}^{\1,\5,\1\0}$ \\ \hline
\end{tabular}
\caption{The sources of the scalar potential in the NMCHM.}\label{tab:sources}
\end{table}

According to the calculability, the sources of the scalar potential are also classified into two types. The first type is the IR contributions, which come from the leading operators in Lagrangians \Eq{spin-1}, \Eq{spin-1/2} and Eqs.~(\ref{6+1}) to (\ref{15+15}). When integrating out the heavy resonances and require suitable Weinberg sum rules, the IR contributions are calculable and expressed in terms of the resonances masses and couplings. The second type, denoted as the UV contributions, are from the local higher dimensional operators which depend on the interactions above the cutoff scale and their interplay with the $SO(6)$-breaking interactions. This type of contributions is incalculable but only estimated by NDA~\cite{Panico:2011pw}. Unfortunately, NDA shows the UV contributions $\gtrsim$ IR contributions~\cite{Marzocca:2012zn}, thus strictly speaking the scalar potential $V(h,\eta)$ is not calculable in CHMs. To ensure the calculability, and partially inspired by the pion mass mechanism in QCD, Ref.~\cite{Marzocca:2012zn} proposes the MHP hypothesis, which assumes the UV contributions are negligible due to some unknown mechanism of the underlying theory. MHP has been generally adopted in the studies of CHMs~\cite{Marzocca:2012zn,Pomarol:2012qf,Redi:2012ha,Marzocca:2014msa,Banerjee:2017qod}. However, as we will demonstrate, under MHP all three NMCHMs we consider fail to trigger SFOEWPT. To realize a SFOEWPT, the UV contributions must be included. A summary of the sources of the scalar potential is given in Table~\ref{tab:sources}.

In the following subsections we will derive the scalar potential for the three benchmark NMCHMs: the $\6+\6$, $\1\5+\6$ and $\1\5+\1\5$ models. For the IR contributions, we express the potential coefficients in terms of the form factor integrals; while for the UV contributions, we list the relevant local operators. For the Higgs field $h$, according to the sources of the $SO(6)$-breaking interactions we can separate the coefficients into
\be
\mu_h^2=\mu_g^2+\mu_f^2,\quad \lambda_h=\lambda_g+\lambda_f,
\ee
where ``$g$'' and ``$f$'' denote the gauge and partial compositeness (fermion) contributions, respectively. The $\mu_\eta^2$ and $\lambda_{\eta,h\eta}$ receive fermion contributions only. We will first discuss the gauge contributions and then the fermion contributions for various embeddings.

\subsection{Contribution from vector bosons}

The gauge contributions are universal for all benchmark NMCHMs. Generally, the gauge-induced potential can be written in a polynomial form
\be\label{Vg}
V_g(h)=\frac{\mu_g^2}{2}h^2+\frac{\lambda_g}{4}h^4,
\ee
where the coefficients receive contributions from both IR and UV sources.

\textbf{The IR contributions}: Integrating out the $\rho$ and $a$ resonances in \Eq{spin-1} we get the Lagrangian involving vector bosons up to quadratic terms in the momentum space
\be\label{EW_form}
\mL_\rho\to\frac12P_T^{\mu\nu}\left(-p^2B_\mu B_\nu-p^2\tr\left[W_\mu W_\nu\right]+\Pi_0(p^2)\tr\left[A_\mu A_\nu\right]+\Pi_1(p^2)\Sigma^\dagger A_\mu A_\nu\Sigma\right),
\ee
where $\Pi_{0,1}(p^2)$ are form factors, and $A_\mu$ is defined \Eq{A_definition}. The transverse and longitudinal projection operators are defined as
\be
P_T^{\mu\nu}=g^{\mu\nu}-\frac{p^\mu p^\nu}{p^2},\quad P_L^{\mu\nu}=\frac{p^\mu p^\nu}{p^2},
\ee
respectively. Under the unitary gauge, \Eq{EW_form} becomes
\begin{multline}\label{H_form_g}
\mL_\rho\to\frac12P_T^{\mu\nu}\Big\{\left(-p^2+\frac{g'^2_0}{g_0^2}\Pi_0(p^2)\right)B_\mu B_\nu+\left(-p^2+\Pi_0(p^2)\right)W_\mu^a W_\nu^a\\
+\frac{\Pi_1(p^2)}{4}\frac{h^2}{f^2}\left[W_\mu^1W_\nu^1+W_\mu^2W_\nu^2+\left(W_\mu^3-\frac{g_0'}{g_0}B_\mu\right)\left(W_\nu^3-\frac{g_0'}{g_0}B_\nu\right)\right]\Big\},
\end{multline}
and it contributes to the Higgs potential as~\cite{Agashe:2004rs}
\be\label{V_g}
V_g^{\rm IR}(h)\approx\frac{6}{2}\int\frac{d^4Q}{(2\pi)^4}\ln\left(1+\frac{\Pi_1}{4\Pi_W}\frac{h^2}{f^2}\right)
+\frac{3}{2}\int\frac{d^4Q}{(2\pi)^4}\ln\left[1+\left(\frac{g'^2_0}{g^2_0}\frac{\Pi_1}{4\Pi_B}+\frac{\Pi_1}{4\Pi_W}\right)\frac{h^2}{f^2}\right],
\ee
where $Q^2\equiv-p^2$, $\Pi_W=Q^2+\Pi_0$ and $\Pi_B=Q^2+(g'^2_0/g_0^2)\Pi_0$. Above result is derived under the assumption of Landau gauge $\xi=0$. Only in this gauge can we omit the contributions from the ghost fields~\cite{Jackiw:1974cv}. An expansion of \Eq{V_g} up to $h^4$ can give a very good approximation, because higher order terms are suppressed by $g_0^2v^2/f^2$. Matching the polynomial potential to \Eq{Vg}, we get
\be\label{g_contribution}\begin{split}
(\mu_g^2)^{\rm IR}=&~\frac{3}{4f^2}\int\frac{d^4Q}{(2\pi)^4}\left(\frac{g'^2_0}{g_0^2}\frac{\Pi_1}{\Pi_B}+3\frac{\Pi_1}{\Pi_W}\right),\\
(\lambda_g)^{\rm IR}=&~-\frac{3}{16f^4}\int\frac{d^4Q}{(2\pi)^4}\left[2\left(\frac{\Pi_1}{\Pi_W^2}\right)^2+\left(\frac{g'^2_0}{g^2_0}\frac{\Pi_1}{\Pi_B}+\frac{\Pi_1}{\Pi_W}\right)^2\right].
\end{split}\ee
Since $\Pi_{B,W}\sim Q^2$ at large momentum, the coefficients are quadratic divergent. To get a convergent $V_g$, the $\Pi_1$ form factor should at least have a scaling $Q^{-4}$. This can be realized via suitable Weinberg sum rules, as we will see in Section~\ref{sec:6+6SFOEWPT}.

\textbf{The UV contributions}: This part of contributions comes from the higher order operators, which can be written down using the spurion trick~\cite{Panico:2011pw}. We rewrite the gauge field as
\be
gA_\mu=gT_L^aW_\mu^a+g'T_R^3B_\mu\equiv\mG_{\bar Aa}T^{\bar A}W_\mu^a+\mG'_{\bar A}T^{\bar A}B_\mu.
\ee
The symmetry of the theory is formally extended to $SO(6)\times SU(2)_0\times U(1)_0$, in which the spurions have quantum number
\be
\mG_{\bar Aa}:(\6,\3_0),\quad \mG'_{\bar A}:(\6,\1_0).
\ee
The VEVs of the spurions
\be
\ave{\mG_{\bar Aa}}=g\delta_{\bar Aa_L},\quad \ave{\mG'_{\bar A}}=g'\delta_{\bar A3_R},
\ee
break the $SO(6)\times SU(2)_0\times U(1)_0$ back to $SU(2)_L\times U(1)_Y\times U(1)_\eta$. The spurions can be used to count the number of gauge insertions when generating a specific operator. Denoting $\mG_a=\mG_{\bar Aa}T^{\bar A}$ and $\mG'=\mG'_{\bar A}T^{\bar A}$, the relevant operators for $h$ potential are
\be
c_gf^4\Sigma^\dagger \mG_a\mG_a\Sigma,\quad c_{g'}f^4\Sigma^\dagger\mG'\mG'\Sigma,\quad \frac{d_g}{16\pi^2}f^4\left(\Sigma^\dagger \mG_a\mG_a\Sigma\right)^2,\quad \frac{d_{g'}}{16\pi^2}f^4\left(\Sigma^\dagger \mG'\mG'\Sigma\right)^2,
\ee
where the coefficients $c_{g,g'}$ and $d_{g,g'}$ are all $\mO(1)$ according to NDA. Matching above operators to \Eq{Vg} yields
\be\label{g_UV}
(\mu_g^2)^{\rm UV}=c_g\frac{3g^2}{2}f^2+c_{g'}\frac{g'^2}{2}f^2,\quad
(\lambda_g)^{\rm UV}=d_g\frac{9g^4}{64\pi^2}+d_{g'}\frac{g'^4}{64\pi^2}.
\ee

\subsection{Contribution from fermions: the $\6+\6$ model}

The fermion-induced potential of all kinds of embeddings can be generally written in
\be\label{Vf}
V_{f}(h,\eta)=\frac{\mu_f^2}{2}h^2+\frac{\lambda_f}{4}h^4+\frac{\mu_\eta^2}{2}\eta^2+\frac{\lambda_\eta}{4}\eta^4+\frac{\lambda_{h\eta}}{2}h^2\eta^2,
\ee
and the contributions to the coefficients can be classified into IR and UV ones.

\textbf{The IR contributions}: Integrating out the top partners in \Eq{6+6}, the general fermion Lagrangian up to quadratic term is 
\begin{multline}\label{6+6_original}
\mL_{\6+\6}\to\bar q_L^\6\slashed{p}\left(\Pi_0^q(p^2)+\Pi_1^q(p^2)\Sigma\Sigma^\dagger\right)q_L^\6+\bar t_R^\6\slashed{p}\left(\Pi_0^t(p^2)+\Pi_1^t(p^2)\Sigma\Sigma^\dagger\right)t_R^\6\\
+\bar q_L^\6\left(M_0^t(p^2)+M_1^t(p^2)\Sigma\Sigma^\dagger\right)t_R^\6+{\rm h.c.}~,
\end{multline}
where $\Pi_{q,t}^{0,1}(p^2)$ and $M_{0,1}^t(p^2)$ are form factors. Above Lagrangian is greatly simplified under the unitary gauge
\begin{multline}\label{6+6_form}
\mL_{\6+\6}\to\bar t_L\slashed{p}\left(\Pi_0^q+\frac{\Pi_1^q}{2}\frac{h^2}{f^2}\right)t_L
+\bar t_R\slashed{p}\left[\Pi_0^t+\Pi_1^t\left(c_\theta^2\frac{\eta^2}{f^2}
+s_\theta^2\left(1-\frac{h^2+\eta^2}{f^2}\right)\right)\right]t_R\\
-\frac{M_1^t}{\sqrt{2}}\frac{h}{f}\left(s_\theta\sqrt{1-\frac{h^2+\eta^2}{f^2}}+i c_\theta\frac{\eta}{f}\right)\bar t_Lt_R+\hc~,
\end{multline}
where we have chosen the unphysical phase $\phi=\pi/2$ in $t_R^\6$. The $\bar b_L b_L$ form factor is accidentally zero because of the $q_L^\6$ embedding. Note that $(h^2+\eta^2)/f^2<1$ by definition (see \Eq{h_redefinition} for details), thus the square root in \Eq{6+6_form} is always well-defined. In the studies involving dark matter, $\theta=\pi/2$ (for which the $t\bar t\eta$ vertex is absent) is chosen to ensure the $\eta\to-\eta$ symmetry and get a stable dark matter candidate~\cite{Frigerio:2012uc,Marzocca:2014msa}. Here we focus on SFOEWPT where the stability of $\eta$ is unimportant, thus allow $\theta$ to be any real number. 

The effective potential caused by \Eq{6+6_form} is
\begin{multline}\label{logarithm}
V_f^{\rm IR}(h,\eta)\approx-2N_c\int\frac{d^4Q}{(2\pi)^4}\Big\{\ln\left(1+\frac{\Pi_1^q}{2\Pi_0^q}\frac{h^2}{f^2}\right)+\ln\left[1+\frac{\Pi_1^t}{\Pi_0^t}\left(s_\theta^2\left(1-\frac{h^2}{f^2}\right)+c_{2\theta}\frac{\eta^2}{f^2}\right)\right]\\
+\ln\left[1+\frac{1}{Q^2}\frac{|M_1^t|^2}{2\Pi_0^q\Pi_0^t}\frac{h^2}{f^2}\left(s_\theta^2\left(1-\frac{h^2}{f^2}\right)+c_{2\theta}\frac{\eta^2}{f^2}\right)\right]\Big\},
\end{multline}
with $N_c=3$ being the QCD color number of the SM quarks. A good approximation can be obtained by truncating the Taylor expansion up to the quartic term of $h$ and $\eta$, as shown in \Eq{Vf}. The coefficients can be expressed in terms of five basic integrals~\cite{Banerjee:2017qod},
\be\label{basic_integrals}
\alpha_{q,t}=\frac{N_c}{f^2}\int\frac{d^4Q}{(2\pi)^4}\frac{\Pi_1^{q,t}}{\Pi_0^{q,t}},\quad\beta_{q,t}=\frac{N_c}{f^4}\int\frac{d^4Q}{(2\pi)^4}\left(\frac{\Pi_1^{q,t}}{\Pi_0^{q,t}}\right)^2,\quad
\epsilon=\frac{N_c}{f^4}\int\frac{d^4Q}{(2\pi)^4}\frac{|M_1^t|^2}{Q^2\Pi_0^q\Pi_0^t},
\ee 
giving
\be\label{IR_6+6}\begin{split}
(\mu_f^2)^{\rm IR}=-2\alpha_q+4s_\theta^2\alpha_t-4s_\theta^4f^2\beta_t-2s_\theta^2f^2\epsilon,&\quad
(\mu_\eta^2)^{\rm IR}=-4c_{2\theta}\alpha_t+4c_{2\theta}s_\theta^2f^2\beta_t,\\
(\lambda_f)^{\rm IR}=\beta_q+4s_\theta^4\beta_t+4s_\theta^2\epsilon,&\quad
(\lambda_\eta)^{\rm IR}=4c_{2\theta}^2\beta_t,\\
(\lambda_{h\eta})^{\rm IR}=-4c_{2\theta}s_\theta^2\beta_t-2c_{2\theta}\epsilon.&
\end{split}\ee
Note that $(\mu_\eta^2)^{\rm IR}$ and $(\lambda_\eta)^{\rm IR}$ are irrelevant to $\alpha_q$ and $\beta_q$ due to the $U(1)_\eta$ symmetry of $q_L^\6$. In addition, provided $\theta=\pi/4$, $\eta$ would decouple from the effective potential. This is because in this limit even $t_R^\6$ has a definite $U(1)_\eta$ quantum number: $-1/\sqrt{2}$, and then $\eta$ is a true NGB that free of potential~\cite{Gripaios:2009pe}. The form factors in the basic integrals of \Eq{basic_integrals} can be derived for the QCD-like underlying theory in terms of the masses of top partners and mixing couplings. Although they are generally divergent, by imposing suitable Weinberg sum rules we can make them converge and get the finite results of \Eq{IR_6+6}. This will be done in Section~\ref{sec:6+6SFOEWPT}.

Before turning to the UV contributions, we demonstrate that SFOEWPT cannot be triggered by the IR contributions alone. The issue is from $(\mu_\eta^2)^{\rm IR}$ and $(\lambda_\eta)^{\rm IR}$. SFOEWPT needs a local minimum along the $\eta$ direction, which is 
\be
w=\sqrt{-\frac{\mu_\eta^2}{\lambda_\eta}}\xrightarrow[]{\rm IR~only}\sqrt{\frac{\alpha_t-s_\theta^2f^2\beta_t}{c_{2\theta}\beta_t}}.
\ee
However, as $\alpha_t$ and $\beta_t$ come from the expansion of the same logarithm, $\abs{\alpha_t}\gg \beta_tf^2$ is expected, thus $w^2|_{\rm IR}\gg f^2$. This inequality can never be achieved because $\eta<f$ is given by its definition, see \Eq{h_redefinition}~\footnote{One may concern that in case $w\gtrsim f$, the perturbative expansion of $\eta^2/f^2$ in the logarithms of \Eq{logarithm} is not valid and we cannot use the polynomial \Eq{V_scalar} to describe the Higgs potential. However, we found that even for $\eta$ comparable with $f$, the statement about the local minimum $w$ remains robust. In Appendix~\ref{app:polynomial} we provide a discussion about this issue.}. Therefore, the IR contributions fail to trigger SFOEWPT.

\textbf{The UV contributions}: A spurion approach is used to rewrite
\be
q_L^\6=\mQ^\6 q_L,\quad t_R^\6=\mT^\6 t_R,
\ee
where the spurions have quantum numbers
\be
\mQ^\6:(\6_{2/3},\2_{-1/6}),\quad \mT^\6:(\6_{2/3},\1_{-2/3}),
\ee
under the extended $SO(6)\times U(1)_X\times SU(2)_0\times U(1)_0$ group. Their VEVs, 
\be
\ave{\mQ^\6}=\frac{1}{\sqrt2}\begin{pmatrix} 0 & 0 & i & -1 & 0 & 0 \\ i & 1 & 0& 0&0&0\end{pmatrix}^T,\quad 
\ave{\mT^\6}=\begin{pmatrix} 0&0&0&0&e^{i \phi } c_\theta & s_\theta \end{pmatrix}^T,
\ee
break the symmetry down to $SU(2)_L\times U(1)_Y$. The operators relevant for scalar potential are
\be\begin{split}
c_f^L\abs{y_L}^2f^4\Sigma^\dagger\mQ^\6\mQ^{\6\dagger}\Sigma,&\quad c_f^R\abs{y_R}^2f^4\Sigma^\dagger\mT^\6\mT^{\6\dagger}\Sigma,\\
\frac{d_f^L}{16\pi^2}\abs{y_L}^4f^4\left(\Sigma^\dagger\mQ^\6\mQ^{\6\dagger}\Sigma\right)^2,&\quad
\frac{d_f^R}{16\pi^2}\abs{y_R}^4f^4\left(\Sigma^\dagger\mT^\6\mT^{\6\dagger}\Sigma\right)^2,
\end{split}\ee
where the coefficients $c_f^{L,R}$ and $d_f^{L,R}$ are the $\mO(1)$ Wilson coefficients according to NDA. The contributions to the scalar potential \Eq{Vf} are then
\be\label{UV_6+6}\begin{split}
(\mu_f^2)^{\rm UV}=&~c_f^L\abs{y_L}^2f^2-2c_f^R\abs{y_R}^2f^2s_\theta^2-\frac{d_f^R}{4\pi^2}\abs{y_R}^4f^2s_\theta^4,\\
(\mu_\eta^2)^{\rm UV}=&~2c_f^R\abs{y_R}^2f^2c_{2\theta}+\frac{d_f^R}{4\pi^2}\abs{y_R}^4f^2s_\theta^2c_{2\theta},\\
(\lambda_f)^{\rm UV}=&~\frac{d_f^L}{16\pi^2}\abs{y_L}^4+\frac{d_f^R}{4\pi^2}\abs{y_R}^4s_\theta^4,\\
(\lambda_\eta)^{\rm UV}=&~\frac{d_f^R}{4\pi^2}\abs{y_R}^4c_{2\theta}^2,\quad(\lambda_{h\eta})^{\rm UV}=-\frac{d_f^R}{4\pi^2}\abs{y_R}^4s_\theta^2c_{2\theta}.
\end{split}\ee
The $c_f^L$ and $d_f^L$ don't contribute to $(\mu_\eta^2)^{\rm UV}$ and $(\lambda_\eta)^{\rm UV}$ because of the $U(1)_\eta$ symmetry. Again, $\eta$ decouples if $\theta=\pi/4$.

The combination of IR and UV contributions gives the complete fermion-induced potential for the $\6+\6$ NMCHM. If $d_f^{R}$ is large enough, then $\lambda_\eta$ receives an enhancement and $w^2\ll f^2$ may be satisfied. In Section~\ref{sec:6+6SFOEWPT} we will show numerically this is indeed the case, i.e. SFOEWPT can be triggered in the $\6+\6$ NMCHM when both IR and UV contributions are taken into account.

\subsection{Contribution from fermions: the $\1\5+\6$ model}

In this subsection, we match the IR and UV contributions of the $\1\5+\6$ model into the general potential in \Eq{Vf}.

\textbf{The IR contributions}: Integrating out the top partners from \Eq{15+6} gives
\begin{multline}\label{L15}
\mL_{\1\5+\6}\to\left(\Pi_0^q(p^2)\tr\left[\bar q_L^{\1\5}\slashed{p}q_L^{\1\5}\right]+\Pi_1^q(p^2)\Sigma^\dagger\bar q_L^{\1\5}\slashed{p}q_L^{\1\5}\Sigma\right)\\+\left(\Pi_0^t(p^2)\bar t_R^\6\slashed{p}t_R^\6+\Pi_1^t(p^2)\bar t_R^\6\slashed{p}\Sigma\Sigma^\dagger t_R^\6\right)
+M_1^t(p^2)\Sigma^\dagger\bar q_L^{\1\5}t_R^\6+\hc~,
\end{multline}
which is simplified in the unitary gauge as 
\begin{multline}\label{15+6_form}
\mL_{\1\5+\6}\to\bar b_L\slashed{p}\left(\Pi_0^q+\frac{\Pi_1^q}{2}\frac{\eta^2}{f^2}\right)b_L+\bar t_L\slashed{p}\left[\Pi_0^q+\frac{\Pi_1^q}{2}\left(\frac{h^2}{2f^2}+\frac{\eta^2}{f^2}\right)\right]t_L
\\+\bar t_R\slashed{p}\left[\Pi_0^t+\Pi_1^t\left(c_\theta^2\frac{\eta^2}{f^2}
+s_\theta^2\left(1-\frac{h^2+\eta^2}{f^2}\right)\right)\right]t_R
-\frac{M_1^t}{2}\frac{h}{f}c_\theta\bar t_Lt_R+{\rm h.c.~},
\end{multline}
where the unphysical phase $\phi$ in $t_R^\6$ is set to $\pi/2$. The corresponding potential can be derived, expanded up to quartic level and matched to the polynomial potential,
\be\label{15+6_IR}\begin{split}
(\mu_f^2)^{\rm IR}=-\alpha_q+4\alpha_ts_\theta^2-4s_\theta^4f^2\beta_t-c_\theta^2f^2\epsilon,&\quad
(\mu_\eta^2)^{\rm IR}=-4\alpha_q-4\alpha_tc_{2\theta}+4c_{2\theta}s_\theta^2\beta_tf^2,\\
(\lambda_f)^{\rm IR}=\frac{\beta_q}{4}+4s_\theta^4\beta_t,&\quad
(\lambda_\eta)^{\rm IR}=2\beta_q+4c_{2\theta}^2\beta_t,\\
(\lambda_{h\eta})^{\rm IR}=\frac{\beta_q}{2}-4c_{2\theta}s_\theta^2\beta_t.&
\end{split}\ee
And the five basic integrals are the same as \Eq{basic_integrals}.

The $\1\5+\6$ model should be the most hopeful one to realize SFOEWPT using only the IR contributions, because both the embeddings $q_L^{\1\5}$ and $t_R^\6$ break $U(1)_\eta$ and then contribute to $\mu_\eta^2$. Therefore, a cancelation may exist in \Eq{15+6_IR} and reduce $(\mu_\eta^2)^{\rm IR}$ to an acceptable value that gives $w^2\ll f^2$. However, the quartic coefficients suffer from another problem. The condition \Eq{degenerate_vacuum} requires $\lambda_{h\eta}^2>\lambda_h\lambda_\eta$. Since $\lambda_f\gg\lambda_g$~\cite{Marzocca:2012zn}, we expect
\begin{multline}
(\lambda_{h\eta}^2)^{\rm IR}-(\lambda_{h})^{\rm IR}(\lambda_{\eta})^{\rm IR}\approx(\lambda_{h\eta}^2)^{\rm IR}-(\lambda_{f})^{\rm IR}(\lambda_{\eta})^{\rm IR}\\=-\frac{1}{4} \beta_q \left[\beta_q+8(1- c_{2 \theta})\beta_t +2  (1+c_{4 \theta}) \beta_t\right]<0,
\end{multline}
where the last inequality holds because $\beta_{q,t}>0$ by definition. Therefore, the necessary condition for SFOEWPT is broken and then IR contributions from the $\1\5+\6$ cannot realize SFOEWPT.

\textbf{The UV contributions}: For $q_L^{\1\5}$, we introduce spurion as 
\be
\left(q_L^{\1\5}\right)_{IJ}=\left(\mQ^{\1\5}\right)_{IJ\alpha}(q_L)_\alpha,\quad \mQ^{\1\5}:(\1\5_{2/3},\2_{-1/6}),
\ee
under the extended $SO(6)\times U(1)_X\times SU(2)_0\times U(1)_0$ group (here $\alpha=1,2$ is the subscript of $SU(2)_0$ elementary representation). The VEV of $\mQ^{\1\5}$ can be inferred from \Eq{qL15} thus not shown here. The relevant operators for the scalar potential are

For the fermion contributions, we have
\be
c_f^L\abs{y_L}^2f^4\Sigma^\dagger\mQ^{\1\5}_{\alpha}\mQ^{\1\5\dagger}_{\alpha}\Sigma,\quad 
\frac{d_f^L}{16\pi^2}\abs{y_L}^4f^4\left(\Sigma^\dagger\mQ^{\1\5}_{\alpha}\mQ^{\1\5\dagger}_{\alpha}\Sigma\right)^2,
\ee
where the coefficients $c_f^{L}$ and $d_f^{L}$ are $\mO(1)$ numbers. The spurion relevant to $t_R^\6$ and the corresponding operators have been introduced in last subsection. Combining them together we get the UV contributions to the scalar potential
\be\begin{split}
(\mu_f^2)^{\rm UV}=&~\frac{c_f^L}{2}\abs{y_L}^2f^2-2c_f^R\abs{y_R}^2f^2s_\theta^2-\frac{d_f^R}{4\pi^2}\abs{y_R}^4f^2s_\theta^4,\\
(\mu_\eta^2)^{\rm UV}=&~2c_f^L\abs{y_L}^2f^2+2c_f^R\abs{y_R}^2f^2c_{2\theta}+\frac{d_f^R}{4\pi^2}\abs{y_R}^4f^2s_\theta^2c_{2\theta},\\
(\lambda_f)^{\rm UV}=&~\frac{d_f^L}{64\pi^2}\abs{y_L}^4+\frac{d_f^R}{4\pi^2}\abs{y_R}^4s_\theta^4,\\
(\lambda_\eta)^{\rm UV}=&~\frac{d_f^L}{4\pi^2}\abs{y_L}^4+\frac{d_f^R}{4\pi^2}\abs{y_R}^4c_{2\theta}^2,\quad(\lambda_{h\eta})^{\rm UV}=\frac{d_f^L}{16\pi^2}\abs{y_L}^4-\frac{d_f^R}{4\pi^2}\abs{y_R}^4s_\theta^2c_{2\theta}.
\end{split}\ee
With the assistance of the UV contributions, $\lambda_{h\eta}$ may be enhanced to be larger than $\sqrt{\lambda_h\lambda_\eta}$ and the necessary conditions of SFOEWPT are achieved.

\subsection{Contribution from fermions: the $\1\5+\1\5$ model}

\textbf{The IR contribution}: integrating out the top partners in \Eq{15+15}, the Lagrangian up to quadratic term is 
\begin{multline}
\mL_{\1\5+\1\5}\to\left(\Pi_0^q(p^2)\tr[\bar q_L^{\1\5}\slashed{p}q_L^{\1\5}]+\Pi_1^q(p^2)\Sigma^\dagger\bar q_L^{\1\5}\slashed{p}q_L^{\1\5}\Sigma\right)\\+\left(\Pi_0^t(p^2)\tr[\bar t_R^{\1\5}\slashed{p}t_R^{\1\5}]+\Pi_1^t(p^2)\Sigma^\dagger\bar t_R^{\1\5}\slashed{p}t_R^{\1\5}\Sigma\right)
+M_1^t(p^2)\Sigma^\dagger\bar q_L^{\1\5}t_R^{\1\5}\Sigma+\hc~,
\end{multline}
Using unitary gauge and choosing the phase $\phi=0$ in $t_R^{\1\5}$, we can simplify the expression as 
\begin{multline}
\mL_{\1\5+\1\5}\to\bar b_L\slashed{p}\left(\Pi_0^q+\frac{\Pi_1^q}{2}\frac{\eta^2}{f^2}\right)b_L\\+\bar t_L\slashed{p}\left[\Pi_0^q+\frac{\Pi_1^q}{2}\left(\frac{h^2}{2f^2}+\frac{\eta^2}{f^2}\right)\right]t_L
+\bar t_R\slashed{p}\left[\Pi_0^t+\frac{\Pi_1^t}{4}\left(2s_\theta^2+(1-3s_\theta^2)\frac{h^2}{f^2}\right)\right]t_R
\\+\frac{M_1^t}{4}\frac{h}{f}\left(\sqrt{2}s_\theta \sqrt{1-\frac{h^2+\eta^2}{f^2}}+i\frac{\eta}{f}c_\theta\right)\bar t_Lt_R+\hc~.
\end{multline}
Matching the corresponding potential to the polynomial form \Eq{Vf} gives the coefficients
\be\label{15+15_IR}\begin{split}
(\mu_f^2)^{\rm IR}=-\alpha_q-\alpha_t(1-3s_\theta^2)+\frac{\beta_tf^2}{2}(1-3s_\theta^2)s_\theta^2-\frac{\epsilon f^2}{2}s_\theta^2,&\quad
(\mu_\eta^2)^{\rm IR}=-4\alpha_q,\\
(\lambda_f)^{\rm IR}=\frac{\beta_q}{4}+\frac{\beta_t}{4}(1-3s_\theta^2)^2+\epsilon s_\theta^2,&\quad
(\lambda_\eta)^{\rm IR}=2\beta_q,\\
(\lambda_{h\eta})^{\rm IR}=\frac{\beta_q}{2}-\frac{1-3s_\theta^2}{4}\epsilon.&
\end{split}\ee
The five basic integrals the same as \Eq{basic_integrals}. The $(\mu_\eta^2)^{\rm IR}$ and $(\lambda_\eta)^{\rm IR}$ are independent of $\alpha_t$ and $\beta_t$ because the embedding $t_R^{\1\5}$ conserves $U(1)_\eta$. It is apparent that \Eq{15+15_IR} cannot trigger SFOEWPT, because it gives a $\eta$-direction local minimum $w^2|_{\rm IR}=2\alpha_q/\beta_q\gg f^2$.

\textbf{The UV contributions}: The spurions for $\mT^{\1\5}$ is
\be
\left(t_R^{\1\5}\right)_{IJ}=\left(\mT^{\1\5}\right)_{IJ}t_R,\quad \mT^{\1\5}:(\1\5_{2/3},\1_{-2/3}),
\ee
under the extended $SO(6)\times U(1)_X\times SU(2)_0\times U(1)_0$ group. $\ave{\mT^{\1\5}}$ can be inferred from \Eq{tR15} thus not shown here. The relevant operators are
\be
c_f^R\abs{y_R}^2f^4\Sigma^\dagger\mT^{\1\5}\mT^{\1\5\dagger}\Sigma,\quad
\frac{d_f^R}{16\pi^2}\abs{y_R}^2f^4\left(\Sigma^\dagger\mT^{\1\5}\mT^{\1\5\dagger}\Sigma\right)^2,
\ee
with $c_f^R$ and $d_f^R$ being $\mO(1)$ parameters. The UV contribution to the scalar potential is
\be\begin{split}
(\mu_f^2)^{\rm UV}=&~\frac{c_f^L}{2}\abs{y_L}^2f^2+\frac{c_f^R}{2}\abs{y_R}^2f^2(1-3s_\theta^2)+\frac{d_f^R}{32\pi^2}\abs{y_R}^4f^2s_\theta^2(1-3s_\theta^2),\\
(\mu_\eta^2)^{\rm UV}=&~2c_f^L\abs{y_L}^2f^2,\\
(\lambda_f)^{\rm UV}=&~\frac{d_f^L}{64\pi^2}\abs{y_L}^4+\frac{d_f^R}{64\pi^2}\abs{y_R}^4(1-3s_\theta^2)^2,\\
(\lambda_\eta)^{\rm UV}=&~\frac{d_f^L}{4\pi^2}\abs{y_L}^4,\quad(\lambda_{h\eta})^{\rm UV}=\frac{d_f^L}{16\pi^2}\abs{y_L}^4.
\end{split}\ee
They can help to reduce $w^2$ and match the requirement of SFOEWPT.

\section{Realizing SFOEWPT in the $\6+\6$ NMCHM}\label{sec:6+6SFOEWPT}

In last section, we have seen the necessary conditions of SFOEWPT are not satisfied by the IR contributions alone, for all benchmark models we considered. However, we also showed that with the help of the UV contributions, SFOEWPT may exist in the $\6+\6$, $\1\5+\6$ and $\1\5+\1\5$ models. In this section, we take the $\6+\6$ NMCHM as an example to investigate the SFOEWPT under the combination of the IR and UV contributions.

\subsection{Calculating the IR contributions with Weinberg sum rules}

The UV contributions to $V(h,\eta)$ of the $\6+\6$ model have been given in Eqs.~(\ref{g_UV}) and (\ref{UV_6+6}), while the IR contributions are expressed as the integrals of the form factors in Eqs.~(\ref{g_contribution}) and (\ref{IR_6+6}). For a QCD-like strong dynamics, the form factors can be written explicitly and the integrals can be evaluated with the help of suitable sum rules.

\textbf{Gauge contributions}: The $\Pi_{0,1}(p^2)$ are expressed by the strong couplings and vector resonances masses~\cite{Contino:2010rs}
\be\label{Pi_definition}\begin{split}
\Pi_0(p^2)=&~g_0^2p^2\sum_{n=1}^{N_\rho}\frac{f_{\rho(n)}^2}{p^2-M_{\rho(n)}^2},\\
\Pi_1(p^2)=&~g_0^2f^2+2g_0^2p^2\left(\sum_{n=1}^{N_a}\frac{f_{a(n)}^2}{p^2-M_{a(n)}^2}-\sum_{n=1}^{N_\rho}\frac{f_{\rho(n)}^2}{p^2-M_{\rho(n)}^2}\right),
\end{split}\ee
where $f_{\rho(n)}\equiv M_{\rho(n)}/g_{\rho(n)}$. Then in \Eq{H_form_g}, the kinetic terms of $B$ and $W$ fields are modified into
\be
-\frac{p^2}{2}P_T^{\mu\nu}\left(1+\sum_{n=1}^{N_\rho}\frac{g'^2_0}{g_{\rho(n)}^2}\right)B_\mu B_\nu,\quad
-\frac{p^2}{2}P_T^{\mu\nu}\left(1+\sum_{n=1}^{N_\rho}\frac{g_0^2}{g_{\rho(n)}^2}\right)W_{\mu}^aW^a_{\nu}.
\ee
A field redefinition is needed to get the canonical kinetic terms, i.e. $W_\mu^a\to(g/g_0)W_\mu^a$ and $B_\mu\to(g'/g'_0)B_\mu$, where
\be\label{g_redefinition}
g^2=g_0^2\left(1+\sum_{n=1}^{N_\rho}\frac{g_0^2}{g_{\rho(n)}^2}\right)^{-1},\quad g^2=g'^2_0\left(1+\sum_{n=1}^{N_\rho}\frac{g'^2_0}{g_{\rho(n)}^2}\right)^{-1},
\ee
are the physical gauge couplings. After this redefinition, the $g_0^{(\prime)}$ in $\Pi_{0,1}$ can be replaced by $g^{(\prime)}$.

As mentioned in previous section, the convergent IR contributions require a scaling $\Pi_1\sim Q^{-4}$. By expanding the second line in \Eq{Pi_definition}, we find that means the following two equations,
\be
\sum_{n=1}^{N_\rho}f_{\rho(n)}^2=\frac{f^2}{2}+\sum_{n=1}^{N_a}f_{a(n)}^2;\quad
\sum_{n=1}^{N_\rho}f_{\rho(n)}^2M_{\rho(n)}^2=\sum_{n=1}^{N_a}f_{a(n)}^2M_{a(n)}^2,
\ee
known as the Weinberg first and second sum rules, which are first proposed in the study of QCD $\rho$-mesons~\cite{Weinberg:1967kj}. If we further assume the lightest resonances dominate, i.e. $N_\rho=N_a=1$, then the sum rules become
\be\label{sum_rules_I}
f_\rho^2=\frac{f^2}{2}+f_a^2,\quad f_\rho^2M_\rho^2=f_a^2M_a^2.
\ee
Note that above equation implies $M_a>M_\rho$. In such a case, the form factors become 
\be
\Pi_0(Q^2)=g^2Q^2\frac{f_{\rho}^2}{Q^2+M_{\rho}^2},\quad
\Pi_1(Q^2)=\frac{g^2f^2M_\rho^2M_a^2}{(Q^2+M_\rho^2)(Q^2+M_a^2)},
\ee
and \Eq{g_contribution} can be calculated analytically. If we impose $\Pi_{B,W}\approx Q^2$ (the error of this approximation is $\mO(g^2/g_\rho^2)$, small enough to neglect), the results are quite simple
\be\label{g_IR}\begin{split}
(\mu_g^2)^{\rm IR}=&~\frac{3(3g^2+g'^2)}{64\pi^2}\frac{M_\rho^2M_a^2}{M_a^2-M_\rho^2}\ln\frac{M_a^2}{M_\rho^2},\\
(\lambda_g)^{\rm IR}=&~\frac{3\left[2g^4+(g^2+g'^2)^2\right]}{256\pi^2(M_a^2-M_\rho^2)^2}\left[M_a^4+\frac{M_\rho^4(M_\rho^2-3M_a^2)}{M_a^2-M_\rho^2}\ln\frac{M_a^2}{M_W^2}+(a\leftrightarrow\rho)\right].
\end{split}\ee
We have used a cutoff $Q^2=M_W^2$ to regularize the IR divergence of $(\lambda_g)^{\rm IR}$~\footnote{Actually the original expression \Eq{V_g} is IR safe, while the IR divergence exists only in the perturbative expansion \Eq{g_contribution}. We have numerically verified that $\lambda_g^{\rm IR}$ is not sensitive to the cutoff we choose.}. Note that $(\mu_g^2)^{\rm IR}$ is positive definite. Comparing Eqs.~(\ref{g_IR}) and (\ref{g_UV}), one can see $(\mu_g^2)^{\rm UV}\gtrsim (\mu_g^2)^{\rm IR}$ because $g_\rho\lesssim4\pi$; while $(\lambda_g)^{\rm UV}\sim (\lambda_g)^{\rm IR}$. Thus in general the UV contribution is not negligible.

\textbf{Fermion contributions}: the fermion form factors $\Pi_{0,1}^{q,t}(p^2)$ and $M_{0,1}^t(p^2)$ are expressed in terms of the resonances masses and coupling constants,
\be
\Pi_0^{q,t}(p^2)=1-\sum_{n=1}^{N_\5}\frac{|y_{L,R}^{\5(n)}|^2f^2}{p^2-M^2_{\5(n)}},\quad \Pi_1^{q,t}(p^2)=\sum_{n=1}^{N_\5}\frac{|y_{L,R}^{\5(n)}|^2f^2}{p^2-M^2_{\5(n)}}-\sum_{n=1}^{N_\1}\frac{|y_{L,R}^{\1(n)}|^2f^2}{p^2-M^2_{\1(n)}},
\ee
and
\be\begin{split}
M_0^t(p^2)=&~-\sum_{n=1}^{N_\5}\frac{y^{\5(n)}_L(y^{\5(n)}_{R})^*f^2M_{\5(n)}}{p^2-M^2_{\5(n)}},\\
M_1^t(p^2)=&~\sum_{n=1}^{N_\5}\frac{y^{\5(n)}_L(y^{\5(n)}_{R})^*f^2M_{\5(n)}}{p^2-M^2_{\5(n)}}-\sum_{n=1}^{N_\1}\frac{y^{\1(n)}_L(y^{\1(n)}_{R})^*f^2M_{\1(n)}}{p^2-M^2_{\1(n)}}.
\end{split}\ee
Defining $Q^2=-p^2$, now the IR-driven coefficients \Eq{IR_6+6} can be evaluated. To converge the integrals, the following scaling is needed,
\be
\Pi_1^{q,t}\sim\frac{1}{Q^6},\quad M_1^t\sim\frac{1}{Q^2}.
\ee
While the second scaling is already satisfied, the first one requires two sets of sum rules 
\be\label{F_sum_rules}
\sum_{n=1}^{N_\5}\left|y_{L,R}^{\5(n)}\right|^2=\sum_{n=1}^{N_\1}\left|y_{L,R}^{\1(n)}\right|^2,\quad 
\sum_{n=1}^{N_\5}\left|y_{L,R}^{\5(n)}\right|^2M_{\5(n)}^2=\sum_{n=1}^{N_\1}\left|y_{L,R}^{\1(n)}\right|^2M_{\1(n)}^2.
\ee

Assuming the lightest resonances dominate, we consider the particle spectrum $N_\5=1$ and $N_\1=2$. In this case \Eq{F_sum_rules} reduces to
\be\label{SR}
\left|y_{L,R}^\5\right|^2=\left|y_{L,R}^\1\right|^2+\left|y_{L,R}^{\1'}\right|^2,\quad 
\left|y_{L,R}^\5\right|^2M_\5^2=\left|y_{L,R}^\1\right|^2M_\1^2+\left|y_{L,R}^{\1'}\right|^2M_{\1'}^2,
\ee
where the heavier singlet top partner is denoted as $\Psi_{\1'}$. The form factors are then
\be\label{Pi_f_1}
\Pi_0^{q,t}(Q^2)=1+\frac{|y_{L,R}^{\5}|^2f^2}{Q^2+M^2_{\5}},
\quad \Pi_1^{q,t}(Q^2)=\frac{\left|y_{L,R}^{\1'}\right|^2f^2\left(M_{\1'}^2-M_\5^2\right)\left(M_{\1'}^2-M_\1^2\right)}{(Q^2+M^2_{\5})(Q^2+M^2_{\1})(Q^2+M^2_{\1'})},
\ee
and
\be\label{Pi_f_2}
M_1^t(Q^2)=\frac{y^{\1}_L(y^{\1}_{R})^*f^2M_{\1}}{Q^2+M^2_{\1}}+\frac{y^{\1'}_L(y^{\1'}_{R})^*f^2M_{\1'}}{Q^2+M^2_{\1'}}-\frac{y^{\5}_L(y^{\5}_{R})^*f^2M_{\5}}{Q^2+M^2_{\5}}.
\ee
The mass of top quark can be read as
\be\label{top_mass}
M_t=\frac{v\abs{s_\theta}}{\sqrt{2}}\frac{M_\5}{\sqrt{M_\5^2+\left|y_L^\5\right|^2f^2}}\frac{M_\5}{\sqrt{M_\5^2+\left|y_R^\5\right|^2f^2}}\left|\frac{y^{\1}_Ly^{\1*}_{R}f}{M_{\1}}+\frac{y^{\1'}_Ly_R^{\1'*}f}{M_{\1'}}-\frac{y^{\5}_Ly^{\5*}_{R}f}{M_{\5}}\right|\sqrt{1-\frac{v^2}{f^2}},
\ee
while the bottom quark remains massless. A mass hierarchy $M_{\1'}>M_{\5}>M_\1$ can be derived from \Eq{SR}. 

Given \Eq{g_IR}, Eqs.~(\ref{Pi_f_1}) and (\ref{Pi_f_2}), the quantitative connection between the IR contributions and resonances masses and couplings are known and the numerical study is in order~\footnote{Ref.~\cite{Marzocca:2014msa} calculates the case $\theta=\pi/2$, where $\eta$ is a dark matter candidate. Here we consider a general $\theta$.}.

\subsection{SFOEWPT and gravitational waves}

\begin{table}
\footnotesize\renewcommand\arraystretch{2}\centering
\begin{tabular}{|c|c|c|}
 \hline
$\6+\6$ & Gauge-induced & Fermion-induced \\ \hline
IR & \tabincell{c}{$\Pi_0(Q^2)=\frac{g^2Q^2f_{\rho}^2}{Q^2+M_{\rho}^2}$,\\ $\Pi_1(Q^2)=\frac{g^2f^2M_\rho^2M_a^2}{(Q^2+M_\rho^2)(Q^2+M_a^2)}$} & \tabincell{c}{$\Pi_0^{q,t}(Q^2)=1+\frac{|y_{L,R}^{\5}|^2f^2}{Q^2+M^2_{\5}}$,\\ $\Pi_1^{q,t}(Q^2)=\frac{\left|y_{L,R}^{\1'}\right|^2f^2\left(M_{\1'}^2-M_\5^2\right)\left(M_{\1'}^2-M_\1^2\right)}{(Q^2+M^2_{\5})(Q^2+M^2_{\1})(Q^2+M^2_{\1'})}$, \\ $M_1^t(Q^2)=\frac{y^{\1}_L(y^{\1}_{R})^*f^2M_{\1}}{Q^2+M^2_{\1}}+\frac{y^{\1'}_L(y^{\1'}_{R})^*f^2M_{\1'}}{Q^2+M^2_{\1'}}-\frac{y^{\5}_L(y^{\5}_{R})^*f^2M_{\5}}{Q^2+M^2_{\5}}$} \\ \hline
UV & \tabincell{c}{$c_gf^4\Sigma^\dagger \mG_a\mG_a\Sigma,~\frac{d_g}{16\pi^2}f^4\left(\Sigma^\dagger \mG_a\mG_a\Sigma\right)^2$,\\ $c_{g'}f^4\Sigma^\dagger\mG'\mG'\Sigma,~\frac{d_{g'}}{16\pi^2}f^4\left(\Sigma^\dagger \mG'\mG'\Sigma\right)^2$} & \tabincell{c}{$c_f^L\abs{y_L}^2f^4\Sigma^\dagger\mQ^\6\mQ^{\6\dagger}\Sigma,~\frac{d_f^L}{16\pi^2}\abs{y_L}^4f^4\left(\Sigma^\dagger\mQ^\6\mQ^{\6\dagger}\Sigma\right)^2,$ \\ $c_f^R\abs{y_R}^2f^4\Sigma^\dagger\mT^\6\mT^{\6\dagger}\Sigma,~\frac{d_f^R}{16\pi^2}\abs{y_R}^4f^4\left(\Sigma^\dagger\mT^\6\mT^{\6\dagger}\Sigma\right)^2$} \\ \hline
\end{tabular}
\caption{The contributions to the scalar potential in the $\6+\6$ NMCHM. Note that this table is a realization of Table~\ref{tab:sources}.}\label{tab:6+6}
\end{table}

The sources of $V(h,\eta)$ are summarized in Table~\ref{tab:6+6}. Combining the IR and UV parts of $\mu_{h,\eta}^2$ and $\lambda_{h,\eta,h\eta}$, we use the {\tt MultiNest} package~\cite{Feroz:2008xx} to find the allowed parameter space by the SM mass spectrum and the conditions for SFOEWPT. For the IR parts, the variables we use in scan are 
\be
\left\{M_\rho,M_a,f,M_\1,M_\5,M_{\1'},y_L^\5,y_R^\5,\theta\right\},
\ee
while $g_{\rho,a}$ and $y_{L,R}^{\1,\1'}$ are derived via the sum rules. The mass ranges are $2\sim7$ TeV for the vector resonances and $1\sim6$ TeV for the fermion resonances, while $f>0.5$ TeV. For the fermion interactions, all the mixing couplings $\abs{y_{L,R}^{\5,\1,\1'}}$ are constrained within 5. For the UV parts, we consider only the fermion-induced operators $c_f^{L,R}$ and $d_f^{L,R}$, requiring the absolute values of the Wilson coefficients to be smaller than 5. The range of mixing angle in embedding $t_R^\6$ is $\abs{\theta}\in[0,\pi/2]$, where the upper limit is due to the fact that $\theta$ exist only as $s_\theta^2$ thus $(\pi-\theta)$ is equivalent to $\theta$. To satisfy the EW and Higgs measurements, we require the derived $M_h=125.09$ GeV~\cite{ATLAS-CONF-2018-031}, $M_Z=91.1876$ GeV~\cite{ALEPH:2005ab}, and the top mass $M_t=172.9\pm0.4$ GeV~\cite{Tanabashi:2018oca}. To really achieve a SFOEWPT, the bubble nucleation condition \Eq{nucleation_condition} should be satisfied, i.e. there should exist a nucleation temperature $T_n$ giving $S_3(T_n)/T_n\sim140$. Numerically, we use the {\tt CosmoTransitions}~\cite{Wainwright:2011kj} package to derive the $O(3)$-symmetric classical bounce solution for $V_T(h,\eta)$ and get $S_3(T)$, and then solve $T_n$. The allowed parameter points distribute almost uniformly in the $M_{\rho,a}$ and $M_{\5,\1,\1'}$ regions we set. We also verify that the for $\mu_{h,\eta}^2$ and $\lambda_{h,h\eta}$, the IR and UV contributions are comparable; while for $\lambda_\eta$ the UV contributions dominate. The allowed $\abs{\theta}$ lies in $1.0\lesssim\abs{\theta}\leqslant\pi/2$, thus a dark matter scenario for $\eta$ (corresponding to $\theta=\pi/2$) is still possible. However, as pointed out in Ref.~\cite{Alanne:2014bra}, under the requirement of SFOEWPT, such a singlet can only contribute a subdominant component after taking into account the direct search bounds.

\begin{figure}
\centering
\subfigure{
\label{fig:Meta_f} 
\includegraphics[scale=0.394]{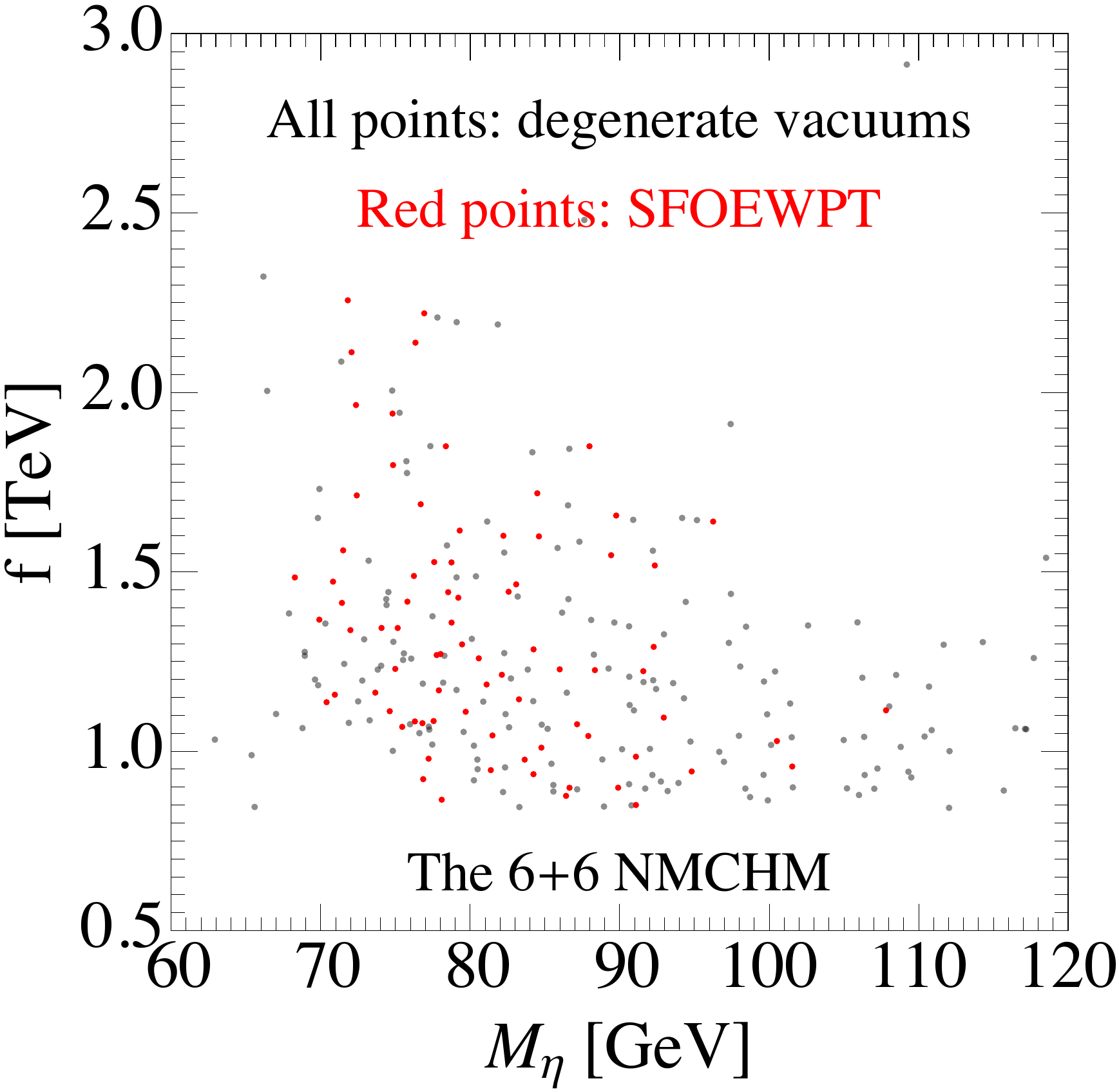}}\quad~~
\subfigure{
\label{fig:Tc_Tn} 
\includegraphics[scale=0.4]{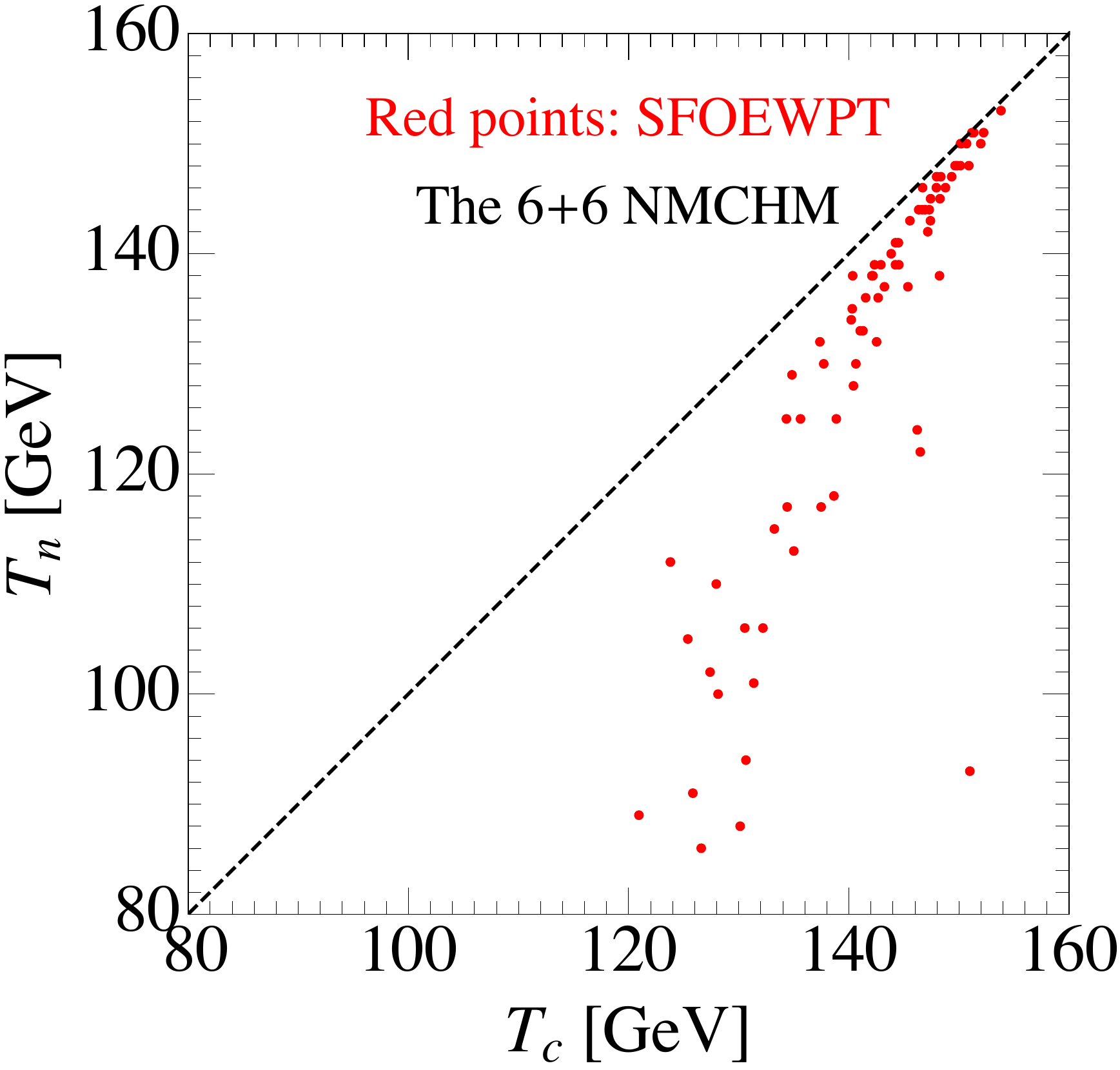}}
\caption{Left: The projections for the parameter points on the $M_\eta-f$ plane. All the points can reproduce SM mass spectrum and give degenerate vacuums at critical temperature $T_c$, while only the red points give SFOEWPT. Right: the $T_n-T_c$ values for the points with successful SFOEWPT.}
\label{fig:SFOEWPT}
\end{figure}

In Fig.~\ref{fig:SFOEWPT}, we project the surviving parameter points into the $f-M_\eta$ and $T_n-T_c$ planes. In the left panel,  all points in the figure reproduce the SM particle spectrum and give degenerate vacuums, while only the red points can trigger SFOEWPT. The mass of $\eta$ is around 100 GeV and the decay constant $f\gtrsim1$ TeV. The right panel of Fig.~\ref{fig:SFOEWPT} shows the critical temperatures $T_c$ and the nucleation temperatures $T_n$ for the parameter points with successful nucleation. One can find $T_n\sim120$ GeV and $T_n\leqslant T_c$ as expected.

SFOEWPT can produce gravitational waves (GWs) in the early universe. After the cosmological redshift, the peak of GW frequencies are typically mille-Hz~\cite{Grojean:2006bp}, in the sensitive region of a broad class of GW detectors, such as LISA~\cite{Audley:2017drz}, Tianqin~\cite{Luo:2015ght}, Taiji~\cite{Hu:2017mde}, BBO~\cite{Crowder:2005nr} or DECIGO (Ultimate DECIGO)~\cite{Kawamura:2011zz,Kawamura:2006up}. As is pointed out in Ref.~\cite{Grojean:2006bp}, the GWs from SFOEWPT can be reduced into a two-parameter problem. The first crucial parameter is $\alpha$, defined by the ratio of the phase transition latent heat to the radiative energy density of the universe in the SFOEWPT period,
\be
\alpha=\frac{\epsilon}{\rho_{\rm rad}},\quad\epsilon=-\Delta V_T+T_n\Delta\frac{\partial V_T}{\partial T}\Big|_{T_n},\quad  \rho_{\rm rad}=\frac{\pi^2}{30}g_*T_n^4,
\ee
where $v_n$ and $g_*$ are respectively the Higgs VEV and the relativistic degrees of freedom at $T_n$, and ``$\Delta$'' denotes the difference between the EW broken and symmetric phases. The second key parameter is $\beta/H_n$, with $\beta^{-1}$ being the time duration of SFOEWPT, and $H_n$ the Hubble constant when SFOEWPT completed,
\be
\beta=\frac{d}{dt}\left(\frac{S_3}{T}\right)\Big|_{t=t_n},\quad
\frac{\beta}{H_n}=T_n\frac{d}{dT}\left(\frac{S_3}{T}\right)\Big|_{T=T_n},
\ee
where $t_n$ is the cosmic time at $T_n$. The smaller $\beta/H_n$ is, the stronger the phase transition is. The signal strength of GWs is described by
\be
\Omega_{\rm GW}(f)=\frac{1}{\rho_c}\frac{d\rho_{\rm GW}}{d\ln f},
\ee
where $\rho_c$ stands for the critical energy density of the universe today. There are three sources of the phase transition GWs: bubble collision, sound waves in the fluid, and the turbulence in plasma. They are all expressed as numerical formulae in terms of $\alpha$ and $\beta/H_n$ in Ref.~\cite{Caprini:2015zlo}. In our scenario, the velocity of the expanding bubble wall is given by the detonation wave formula~\cite{Steinhardt:1981ct}
\be
v_{\rm w}=\frac{1}{1+\alpha}\left(\frac{1}{\sqrt{3}}+\sqrt{\alpha^2+\frac{2}{3}\alpha}\right),
\ee
and the dominant source of GWs comes from sound waves, while the turbulence is subdominant and the bubble collision contribution is negligible~\cite{Caprini:2015zlo}~\footnote{This conclusion might be modified because the sound wave period has to be appropriately cut when the plasma flow becomes nonlinear. The turbulence may get a lot of remaining kinetic energy and contributes a much stronger signal~\cite{Ellis:2018mja}.}. 

\begin{figure}
\centering
\subfigure{
\label{fig:alpha_beta} 
\includegraphics[scale=0.4]{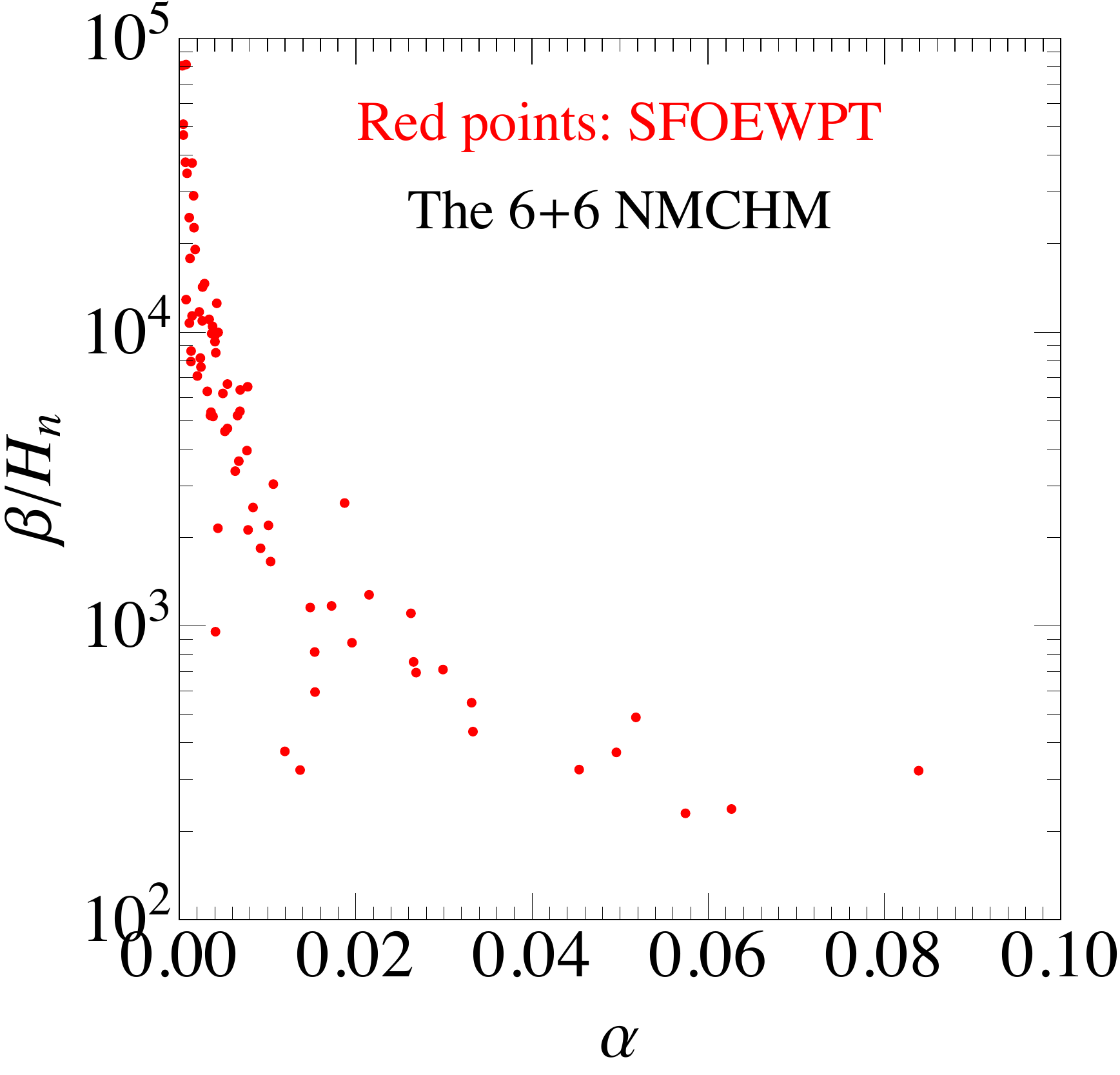}}\quad~~
\subfigure{
\label{fig:GW} 
\includegraphics[scale=0.4]{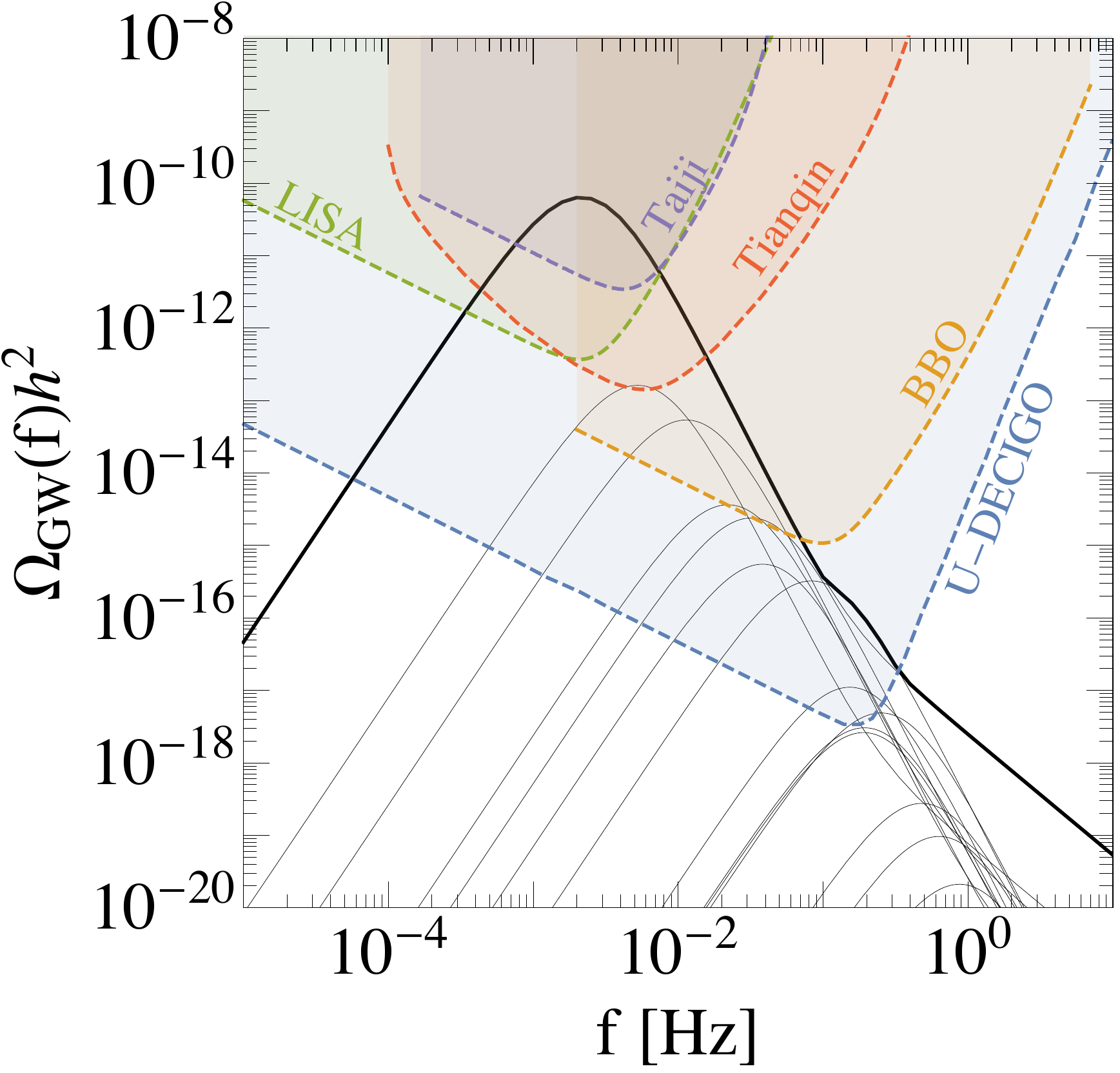}}
\caption{Left: the $\alpha$ and $\beta/H_n$ distribution for the parameter points with SFOEWPT. Right: the GW signals, where the thin black lines are typical GW curves from the data points of the left panel, while the thick black line represents the envelope of all data points.}
\label{fig:GWs}
\end{figure}

For our study, the relativistic degrees of freedom during SFOEWPT is $g_*=106.75+1$, i.e the number of SM plus one real singlet. For the data points with successful nucleation in $\6+\6$ NMCHM, we calculate $\alpha$ and $\beta/H_n$ with {\tt CosmoTransitions} and a homemade codes plugin. The obtained values of $\alpha$ and $\beta/H_n$ are projected in the left panel of Fig.~\ref{fig:GWs}. Using the formulae in~\cite{Caprini:2015zlo} we are able to calculate the GW signal strengths. The results are presented in the right panel of Fig.~\ref{fig:GWs}, where some typical signal curves are plotted in thin black lines while the envelope of all allowed data points are plotted in a thick black line. One can clearly see that GW signals are testable for most future detectors.

\subsection{Collider phenomenology}

The NMCHM is rather predictive and they have very rich phenomenology at the LHC. On one hand, the deviations of the Higgs couplings or oblique parameters can be probed in the EW and Higgs precision measurements~\cite{Banerjee:2017wmg,Franzosi:2016aoo,Niehoff:2016zso}; on the other hand, the composite resonances can be directly discovered~\cite{Franzosi:2016aoo,Niehoff:2016zso}. While no excess is obtained, the experiments have been putting stronger and stronger constrains on the model. 

The discovery of composite resonances would be the smoking gun of the CHMs. In NMCHM, the vector mass terms in \Eq{spin-1} induce EFT operators
\be\label{spin-1_EFT}
\mL_\rho\supset\frac{M_{\rho}^2}{2g_{\rho}^2}\left[\left(g_\rho\rho_{L\mu}^{a}-g_0W_\mu^{a}+\frac{i}{2f^2}H^\dagger\sigma^{a}\Dfbd H\right)^2+\left(g_\rho\rho_{R\mu}^{3}-g_0'B_\mu+\frac{i}{2f^2}H^\dagger\Dfbd H\right)^2\right],
\ee
implying the mixings $\rho_L^a-W^a$ and $\rho_R^3-B$ before EWSB, and the mixing angles are $\approx g/g_\rho$ and $g'/g_\rho$ respectively. As a result, the $\rho_L^{\pm,0}$ and $\rho_R^3~(=\rho_R^0)$ interact with light quarks with couplings $\approx g^2/g_\rho$ and $g'^2/g_\rho$ respectively and can be produced via Drell-Yan process at the LHC and then decay to the SM di-boson channels ($W^\pm Z/W^\pm h$ and $W^+W^-/Zh$). Other vector resonances such as $\rho_R^{\pm}$, $\rho_D^{\pm,0,0*}$, $a_D^{\pm,0,0*}$ and $a_S^0$ interact with quark partons only after EWSB thus the couplings are suppressed by $v^2/f^2$. Therefore, it is hard to probe them via the quark-antiquark fusion at the LHC. The $\rho_R^\pm$ may be produced via vector boson fusion as well, however the cross section is tiny due to the phase space suppression. In summary, the most hopeful channel to probe the vector resonances of NMCHM is the Drell-Yan produced $\rho_L^{\pm,0}$ and $\rho_R^0$. The dominant decay channels of $\rho_L^{\pm,0}$ and $\rho_R^0$ depend on the relation between the masses of vector and fermion resonances~\cite{Liu:2018hum}~\footnote{For $\rho_L^{\pm,0}$ and $\rho_R^0$, the phenomenology is very similar to the case of the $\5+\5$ $SO(5)/SO(4)$ CHM (i.e. the MCHM). Therefore, the corresponding discussions in Ref.~\cite{Liu:2018hum} also apply to here.}. In the region $M_\rho<M_\5$, the SM di-boson channels dominate. While if $M_\5<M_\rho<2M_\5$, the ``heavy-light'' decay modes with $t/b$ plus a top partner (such as $t\bar\Psi_\5$) kinematically open and acquire considerable branching ratios. Although di-boson channels are sub-leading here, they play a important role in phenomenology because of the accurate measurement at the LHC, see below. Finally, if $M_\rho>2M_\5$, the decay modes $\rho_L^{\pm,0}$, $\rho_R^0\to \Psi_\5\bar\Psi_\5$ (so-called ``heavy-heavy'' channels) induced by the interactions in the strong sector
\be
\mL_{\rho\Psi}=c_\rho\bar\Psi_{\5}\gamma^\mu t^{\bar A}\Psi_{\5}(g_{\rho}\rho^{\bar A}_\mu-e^{\bar A}_\mu),\quad c_\rho\sim\mO(1),
\ee
contribute almost 100\% branching ratio because of the large $g_\rho$.

\begin{figure}
\centering
\includegraphics[scale=0.4]{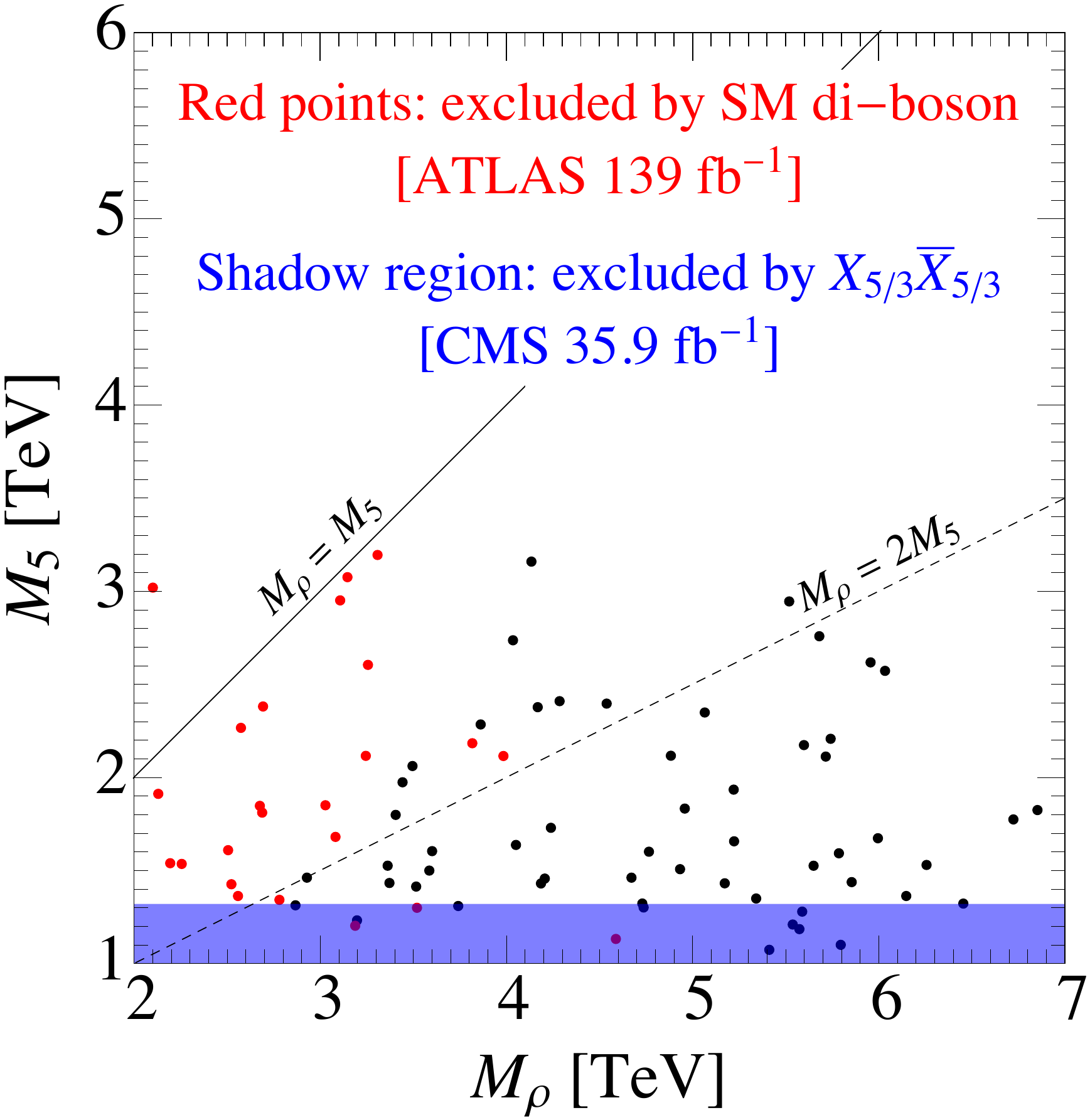}
\caption{The collider phenomenology of the parameter points with SFOEWPT. The SM di-boson bound comes from Ref.~\cite{Aad:2019fbh}, while the $X_{5/3}\bar X_{5/3}$ bound is taken from Ref.~\cite{Sirunyan:2018yun}.}
\label{fig:collider}
\end{figure}

For the $\6+\6$ NMCHM, we found that almost all parameter points yielding SFOEWPT lie in the mass region $M_\rho>M_\5$, as shown in Fig.~\ref{fig:collider}. By recasting Ref.~\cite{Aad:2019fbh} we find most points in region $M_\1<M_\rho<2M_\5$ have been excluded by the SM di-boson searches of ATLAS at 139 fb$^{-1}$, see the red points in Fig.~\ref{fig:collider}. For the region $M_\rho>2M_\5$, the bound is weak that a $\rho$-resonances of $\sim3$ TeV is still allowed. That is because a dedicated search for the heavy-heavy channels is still lacking (see Ref.~\cite{Kim:2019rhy} for a summary of the performed searches after LHC Run II). As in this region the $\rho$-resonances are typically rather broad that $\Gamma_\rho/M_\rho$ can reach $\sim50\%$, the search is more challenging. In this case, the same-sign lepton final state which doesn't require the reconstruction of a resonant peak may be useful to hunt the signal. The studies in Refs.~\cite{Liu:2018hum,Barducci:2015vyf,Vignaroli:2014bpa} show that for 
\be
pp\to\rho_L^\pm\to X_{5/3}\bar X_{2/3}+{\rm c.c.}\to \ell^\pm\ell^\pm+{\rm jets},
\ee
the HL-LHC can reach a region of $M_\rho\gtrsim4.5$ TeV for a $g_\rho\sim2$, 3.

At the LHC, the top partners can be either pair produced via QCD or singly produced via EW interactions. The QCD production is model-independent but suffers from the double-suppression in phase space; while the EW production can probe higher mass scale but depends on the details of the fermion embedding. For the $\6+\6$ NMCHM, matching \Eq{6+6} to SM EFT yields
\be
\mL_{\6+\6}\supset
-s_\theta y_R^\5\left[\bar t_R\left(H^\dagger Q_X-\widetilde H^\dagger Q\right)+\bar t_R\widetilde T\pi_5\right]
-y_L^\1 \bar q_L\widetilde HT_S-ic_\theta y_R^\1  \bar t_RT_S\pi_5.
\ee
According to the Goldstone equivalence theorem, the decay channels of the top partners are
\be
X_{5/3}\to tW^+,\quad X_{2/3},T\to tZ,th,\quad B\to tW^-,\quad \widetilde T,T_S\to bW^+,tZ,th,t\eta.
\ee
The EW fusion production mechanism can also be read, e.g. $tW\to X_{5/3}$. Except for the $T,T_S\to t\eta$ channel, the phenomenology of top partners in the $\6+\6$ NMCHM is quite similar to the case of the $\5+\5$ MCHM and we refer the readers to the relevant study in~\cite{Liu:2018hum} and the references therein. A search based on the CMS 35.9 fb$^{-1}$ data in the $X_{5/3}\bar X_{5/3}$ channel sets a bound $M_\5\geqslant1.32$ TeV, assuming $\Br(X_{5/3}\to tW^+)=100\%$~\cite{Sirunyan:2018yun}. We plot this bound into Fig.~\ref{fig:collider} as the blue shadow region. Note that for a NMCHM based on the bottom-up model building of a specific underlying theory, there may exist extra scalars which are not described in our top-down CCWZ approach. For instance, the NMCHM based on coset $SU(4)/Sp(4)$ and gauge group $Sp(2N_{\rm HC})$ (where $N_{\rm HC}$ is the number of the ``hyper-color'' in the strong sector) contains a real color octet $\pi_8$, a complex color sextet $\pi_6$, and two real gauge singlets $\sigma$ and $\sigma_c$~\cite{Cacciapaglia:2015eqa,Belyaev:2016ftv}. Some of those extra particles are expected to be light and may be decayed from the top partners~\cite{Serra:2015xfa,Bizot:2018tds}, which would even make the bounds weaker because there are no specific searches for those channels yet. A recent study shows that in the same-sign lepton channel, the results from the search for pair-produced $X_{5/3}\to tW^+$ are robust as well for many other exotic decays such as $X_{5/3}\to \bar b\pi_6$, and the HL-LHC can reach a mass of $M_\5\sim1.6$ TeV~\cite{Xie:2019gya}. Another fresh paper~\cite{Cacciapaglia:2019zmj} gives the result for the decay $T\to t\sigma$. For other non-standard decay channels, more detailed studies are still needed.

In short, we've found a lot of parameter points with SFOEWPT in the range $M_{\rho,a}\in[2,7]$ TeV and $M_{\5,\1,\1'}\in[1,6]$ TeV, and a considerable fraction of them lie in $M_\rho>2M_\5$, a region that has been constrained very weakly so far. In this case, current bounds for the vector and fermion resonances are $\sim 3$ TeV and $\sim1.3$ TeV respectively. Therefore, there is plenty of rooms for future LHC experiments to explore this scenario. The excess at the collider can be a good crosscheck of the signals from the GW detectors.

\section{Conclusion}\label{sec:conclusion}

In this article, we study the SFOEWPT scenario in the NMCHM. Within the framework of gauge-invariant thermal corrections to the scalar potential, the SFOEWPT is realized via a two-step phase transition. We have considered various fermion embeddings: for the left-handed doublet $q_L=(t_L,b_L)^T$, we consider \6 or \1\5; while for the right-handed $t_R$, we consider \1, \6 and \1\5. Among the six different combinations, the $\1\5+\1$ model fails to give a massive top quark, while the $\6+\1$ and $\6+\1\5$ models are unable to generate a potential for the singlet $\eta$ thus cannot trigger a two-step phase transition. We then investigate the remained three models $\6+\6$, $\1\5+\6$ and $\1\5+\1\5$ in detail.

We show that if only the IR contributions to the scalar potential are considered, all those three models cannot trigger SFOEWPT. For the $\6+\6$ and $\1\5+\1\5$ models, the problem is $\ave{\eta}\gg f$, which breaks the perturbativity of the theory; while for the $\1\5+\6$ model, the issue is $\lambda_{h\eta}^2>\lambda_h\lambda_\eta$ cannot be satisfied. That means the generally-adopted assumption called minimal Higgs potential hypothesis (MHP), which assumes the IR contributions dominate the UV ones, is incompatible with the SFOEWPT in NMCHM for fermion embeddings up to \1\5. We also demonstrate that the SFOEWPT is hopeful to happen when the UV contributions are added. Taking the $\6+\6$ model a concrete example, we combine the IR and UV contributions and numerically derive its allowed parameter space for SFOEWPT. The GWs from the phase transition are within the sensitive region of the future GW detectors. In addition, the model can be explored at the LHC via the searches for the composite resonances.

At last, we note the conundrum on the wall velocity for EW baryogenesis (EWB) and the GW: a significant GW signal prediction from the EWPT requires supersonic bubble wall expansion velocities $v_{\rm w}$, but the EWB prefers subsonic wall velocities for the effective diffusion. 
At present,  this conundrum is still an open problem. Refs.~\cite{No:2011fi,Alves:2018oct,Alves:2018jsw} suggest that the relevant velocity for baryogenesis is actually not $v_{\rm w}$ but $v_+$, i.e. the relative velocity between the expanding bubble wall and the plasma just in front of it. Hydrodynamics analysis of the plasma shows that it is possible to have supersonic $v_{\rm w}$ but sufficiently low $v_+$~\cite{No:2011fi,Alves:2018oct,Alves:2018jsw} and hence the EWB still works. We leave the detailed study on this issue for the future work.

\acknowledgments

We are grateful to Lian-Tao Wang for the useful discussions about the UV contributions and EW sphaleron. We thank Avik Banerjee, Giacomo Cacciapaglia, Fa Peng Huang, Andrew Long, Michele Redi, Jing Shu and Yang Zhang for the discussions about phase transition and/or composite Higgs model. We are grateful to Guy D. Moore and Michael J. Ramsey-Musolf for discussions on baryon number preservation criterion and phase transition strength. LGB is supported by the National Natural Science Foundation of China under grant No.11605016 and No.11647307. YCW is supported by the Natural Sciences and Engineering Research Council of Canada (NSERC). KPX is supported by Grant Korea NRF 2015R1A4A1042542 and NRF 2017R1D1A1B03030820. KPX would like to express a special thanks to the Mainz Institute for Theoretical Physics (MITP) of the DFG Cluster of Excellence PRISMA$^+$ (Project ID 39083149) for its hospitality, because Appendix~\ref{app:polynomial} of this paper is a consequence of the discussions at MITP.

\appendix
\section{The CCWZ formulae for NMCHM}\label{app:CCWZ}

\textbf{The generators of $SO(6)$ group}: they can be written as~\cite{Niehoff:2016zso}:
\be\begin{split}
[T^{a}_L]_{IJ}=&~-\frac{i}{2}\left[\frac{1}{2}\epsilon^{abc}(\delta_{bI}\delta_{cJ}-\delta_{bJ}\delta_{cI})+(\delta_{aI}\delta_{4J}-\delta_{aJ}\delta_{4I})\right],\\
[T^{a}_R]_{IJ}=&~-\frac{i}{2}\left[\frac{1}{2}\epsilon^{abc}(\delta_{bI}\delta_{cJ}-\delta_{bJ}\delta_{cI})-(\delta_{aI}\delta_{4J}-\delta_{aJ}\delta_{4I})\right],\\
[\hat T^{i}_1]_{IJ}=&~-\frac{i}{\sqrt2}(\delta_{iI}\delta_{5J}-\delta_{iJ}\delta_{5I}),\\
[\hat T^{r}_2]_{IJ}=&~-\frac{i}{\sqrt2}(\delta_{rI}\delta_{6J}-\delta_{rJ}\delta_{6I}),
\end{split}\ee
where ($a=1,2,3$), ($i=1,\cdots,4$), ($r=1,\cdots,5$) and ($I,J=1,\cdots,6$). The normalization is $\tr[T^AT^B]=\delta^{AB}$. We denote the 10 unbroken generators of $SO(5)$ as $T^{\bar A}=\{T_L^a,T_R^a,\hat T_1^i\}$, in which the $\{T_L^a,T_R^a\}$ belong to the $SO(4)\cong SU(2)_L\times SU(2)_R$ subgroup. 

\textbf{The $d$ and $e$ symbols in unitary gauge}: the $d$ symbol is  
\be\begin{split}
d^1_\mu=&~\frac{g_0W_\mu^1}{\sqrt{2}}\frac{h}{f},\quad d^2_\mu=\frac{g_0W_\mu^2}{\sqrt{2}}\frac{h}{f},\quad 
d^3_\mu=\frac{g_0W_\mu^3-g'_0B_\mu}{\sqrt{2}}\frac{h}{f},\\
d^4_\mu=&~\frac{\sqrt{2}}{f}\frac{1}{h^2+\eta^2}\left[\eta\left(h\partial_\mu\eta-\eta\partial_\mu h\right)-\frac{h\left(h\partial_\mu h+\eta\partial_\mu\eta\right)}{\sqrt{1-(h^2+\eta^2)/f^2}}\right],\\
d^5_\mu=&~\frac{\sqrt{2}}{f}\frac{1}{h^2+\eta^2}\left[h\left(\eta\partial_\mu h-h\partial_\mu \eta\right)-\frac{\eta\left(h\partial_\mu h+\eta\partial_\mu\eta\right)}{\sqrt{1-(h^2+\eta^2)/f^2}}\right].
\end{split}\ee
The Goldstone kinetic term is given in \Eq{NGB_kinetic}.

The $e$ symbol has 10 components: $e^{\bar A}_{\mu}=\{e_{L\mu}^{a},e_{R\mu}^{a},e_{1\mu}^{i}\}$, corresponding to the $SO(5)\to SO(4)$ decomposition $\1\0\to(\3,\1)\oplus(\1,\3)\oplus(\2,\2)$. For the $(\3,\1)$ subset, we get
\be\begin{split}
e_{L\mu}^{1}=&~g_0W^1_\mu-\frac{1}{2}g_0W^1_\mu\frac{h^2}{f^2}\left(\frac{1}{1+\sqrt{1-(h^2+\eta^2)/f^2}}\right),\\
e_{L\mu}^{2}=&~g_0W^2_\mu-\frac{1}{2}g_0W^2_\mu\frac{h^2}{f^2}\left(\frac{1}{1+\sqrt{1-(h^2+\eta^2)/f^2}}\right),\\
e_{L\mu}^{3}=&~g_0W^3_\mu-\frac{1}{2}\left(g_0W^3_\mu-g'_0B_\mu\right)\frac{h^2}{f^2}\left(\frac{1}{1+\sqrt{1-(h^2+\eta^2)/f^2}}\right);
\end{split}\ee
while for the $(\1,\3)$ subset we get 
\be\begin{split}
e_{R\mu}^{1}=&~\frac{1}{2}g_0W^1_\mu\frac{h^2}{f^2}\left(\frac{1}{1+\sqrt{1-(h^2+\eta^2)/f^2}}\right),\\
e_{R\mu}^{2}=&~\frac{1}{2}g_0W^2_\mu\frac{h^2}{f^2}\left(\frac{1}{1+\sqrt{1-(h^2+\eta^2)/f^2}}\right),\\
e_{R\mu}^{3}=&~g'_0B_\mu+\frac{1}{2}\left(g_0W^3_\mu-g'_0B_\mu\right)\frac{h^2}{f^2}\left(\frac{1}{1+\sqrt{1-(h^2+\eta^2)/f^2}}\right);
\end{split}\ee
finally the $(\2,\2)$ gives
\be\begin{split}
e_{1\mu}^{1}=&~-\frac{1}{\sqrt{2}}g_0W^1_\mu\frac{h\eta}{f^2}\left(\frac{1}{1+\sqrt{1-(h^2+\eta^2)/f^2}}\right),\\
e_{1\mu}^{2}=&~-\frac{1}{\sqrt{2}}g_0W^2_\mu\frac{h\eta}{f^2}\left(\frac{1}{1+\sqrt{1-(h^2+\eta^2)/f^2}}\right),\\
e_{1\mu}^{3}=&~-\frac{1}{\sqrt{2}}\left(g_0W^3_\mu-g'_0B_\mu\right)\frac{h\eta}{f^2}\left(\frac{1}{1+\sqrt{1-(h^2+\eta^2)/f^2}}\right),\\
e_{1\mu}^{4}=&~\sqrt{2}\frac{\eta\partial_\mu h-h\partial_\mu\eta}{f^2}\left(\frac{1}{1+\sqrt{1-(h^2+\eta^2)/f^2}}\right).
\end{split}\ee

\textbf{The composite resonances}: for the vector resonances, the full expressions of the decompositions in \Eq{vector_decomposition} are 
\be\begin{split}
\rho_{L\mu}^{\pm}=&~\frac{\rho_{L\mu}^1\mp i\rho_{L\mu}^2}{\sqrt{2}},\quad \rho_{L\mu}^{0}=\rho_{L\mu}^3;\quad
\rho_{R\mu}^{\pm}=\frac{\rho_{R\mu}^1\mp i\rho_{R\mu}^2}{\sqrt{2}},\quad \rho_{R\mu}^{0}=\rho_{R\mu}^3;\\
\rho_{D\mu}=&~\begin{pmatrix}\rho_{D\mu}^+\\ \rho_{D\mu}^0\end{pmatrix}=\frac{1}{\sqrt{2}}\begin{pmatrix}\rho_{1\mu}^2+i\rho_{1\mu}^1\\ \rho_{1\mu}^4-i\rho_{1\mu}^3\end{pmatrix},
\end{split}\ee
for the $\rho$-resonances, and
\be
a_{D\mu}=\begin{pmatrix}a_{D\mu}^+\\ a_{D\mu}^0\end{pmatrix}=\frac{1}{\sqrt{2}}\begin{pmatrix}a_{\mu}^2+ia_{\mu}^1\\ a_{\mu}^4-ia_{\mu}^3\end{pmatrix},\quad a_{S\mu}=a_\mu^5.
\ee
for the $a$-resonances. In total, we have 4 singly charged and 7 real neutral vector resonances, in total 15 degrees of freedom.

The decomposition of the top partners are listed in \Eq{fermion_decomposition}. For $\Psi_\5$, the result is
\be
\Psi_\5 = \frac{1}{\sqrt{2}}\begin{pmatrix}
i B - i X_{5/3} &
 B + X_{5/3} &
i T + i X_{2/3} &
 -T + X_{2/3} &
 \widetilde T \end{pmatrix}^T,
\ee
in which we can form two $SU(2)_L\times U(1)_Y$ doublets
\be
Q_X=\begin{pmatrix}X_{5/3}\\ X_{2/3}\end{pmatrix}_{7/6},\quad Q=\begin{pmatrix}T\\ B\end{pmatrix}_{1/6},
\ee
and one singlet $\widetilde T:\1_{2/3}$. The decomposition of $\Psi_{\1\0}$ is a little bit complicated,
\be
\Psi_{\1\0}=t_L^a Y^a+t_R^a K^a+\hat t_1^iJ^i,
\ee
where $[t_{L,R}^a]_{rs}\equiv [T_{L,R}^a]_{rs}$, $[\hat t_1^i]_{rs}\equiv [\hat T_1^i]_{rs}$ with $(r,s=1,...,5)$, and $Y^a$, $K^a$ and $J^i$ are respectively $(\3,\1)_{2/3}$, $(\1,\3)_{2/3}$ and $(\2,\2)_{2/3}$ in $SO(4)\times U(1)_X$. Their explicit expressions are
\be\label{YKJ}{\footnotesize
Y^a=\frac{1}{\sqrt2}\begin{pmatrix}Y_{5/3}+Y_{-1/3}\\ iY_{5/3}-iY_{-1/3}\\ \sqrt{2}Y_{2/3}\end{pmatrix},\quad 
K^a=\frac{1}{\sqrt2}\begin{pmatrix}K_{5/3}+K_{-1/3}\\ iK_{5/3}-iK_{-1/3}\\ \sqrt{2}K_{2/3}\end{pmatrix},\quad 
J^i=\frac{1}{\sqrt{2}}\begin{pmatrix}
i J_{-1/3}- i J_{5/3} \\  J_{-1/3} + J_{5/3} \\ i J_{2/3A} + i J_{2/3B} \\  -J_{2/3A} + J_{2/3B}\end{pmatrix}
},
\ee
where the subscripts denote the electric charges. The $J$ can be further organized into two $SU(2)_L\times U(1)_Y$ doublets
\be
J_X=\begin{pmatrix}J_{5/3}\\ J_{2/3B}\end{pmatrix}_{7/6},\quad J_Q=\begin{pmatrix}J_{2/3A}\\ J_{-1/3}\end{pmatrix}_{1/6}.
\ee

\section{$Zb_L\bar b_L$ coupling in the $q_L^{\1\5}$ embedding}\label{app:Zbb}

The Lagrangian of top partner \Eq{spin-1/2} can be matched to the SM EFT form, yielding 
\be
\mL_\Psi \supset\tr\left[\bar\Psi_{\1\0}\left(i\slashed{\nabla}+g_0'\frac{2}{3}\slashed{B}\right)\Psi_{\1\0}\right]\supset\bar Yi\slashed{D}Y+\bar K_{-1/3}i\slashed{D}K_{-1/3}+\bar J_Qi\slashed{D}J_Q,
\ee
where $D_\mu$ denotes the SM gauge covariant derivative. For a fermion with $SU(2)_L$ quantum number $T_L^3$ and electric charge $Q$, the tree level coupling to $Z$ boson is 
\be
\frac{g}{c_W}\left(T_L^3-s_W^2Q\right).
\ee
For the charge $-1/3$ particles we get 
\be
T_L^3(b_L)=T_L^3(B)=T_L^3(J_{-1/3})=-\frac12,\quad T_L^3(Y_{-1/3})=-1,\quad T_L^3(K_{-1/3})=0.
\ee
Thus the mixing between $b_L$ and $Y$, $K$ will change the $Zb_L\bar b_L$ coupling~\footnote{The $b_L-J_Q$ mixing is safe because of the $P_{LR}$ symmetry~\cite{Agashe:2006at}.}. While for the partial compositeness interactions, 
\be
(\bar{q}_L^{\1\5_A})_{IJ}U_{Jr}\Psi_{\1\0}^{rs}[U^\dagger]_{sI}\supset i\frac{1}{2\sqrt{2}f^2}\bar q_L\sigma^a\widetilde H Y^a\pi^5
-i\frac{1}{2f^2}\bar q_LHK_{-1/3}\pi^5,
\ee
and
\begin{multline}
(\bar{q}_L^{\1\5_B})_{IJ}U_{Jr}\Psi_{\1\0}^{rs}[U^\dagger]_{sI}\supset\\
\frac{1}{\sqrt2 f}\bar q_L\sigma^a\widetilde HY^a\left(1-\frac{2|H|^2+\pi_5^2}{6f^2}\right)-\frac{1}{f}\bar q_LHK_{-1/3}\left(1-\frac{2|H|^2+\pi_5^2}{6f^2}\right).
\end{multline}
There is no $b_L-Y$ or $b_L-K$ mixing for the $\1\5_A$ embedding, as long as $\ave{\pi_5}=\ave{\eta}=0$ at zero temperature. In contrast, those mixings are present in $\1\5_B$. Since the $Zb_L\bar b_L$ couplings are stringently constrained by LEP~\cite{ALEPH:2005ab,Gori:2015nqa}, $\1\5_B$ is strongly disfavored by the experiment and we will not use it to build the NMCHM in this paper.

\section{The validity of polynomial approximation}\label{app:polynomial}

Let's take the $\6+\6$ model as an example. For the IR-driven potential, we expand the logarithms up to quartic level to get a polynomial, see \Eq{V_g} for the gauge part and \Eq{logarithm} for the fermion part, respectively. For the gauge part, the polynomial approximation is valid even for $h\sim f$, because the expansion is suppressed by an additional factor of $g^2/g_\rho^2\ll1$. However, for the fermion-induced potential, the corresponding factor is $|y_{L,R}^{\5,\1,\1'}|^2f^2/M_{\5,\1,\1'}^2$, which can be $\mO(1)$. Therefore, one may concern about the validity of the polynomial approximation when $\eta\sim f$.

\begin{figure}
\centering
\subfigure{
\label{fig:error} 
\includegraphics[scale=0.4]{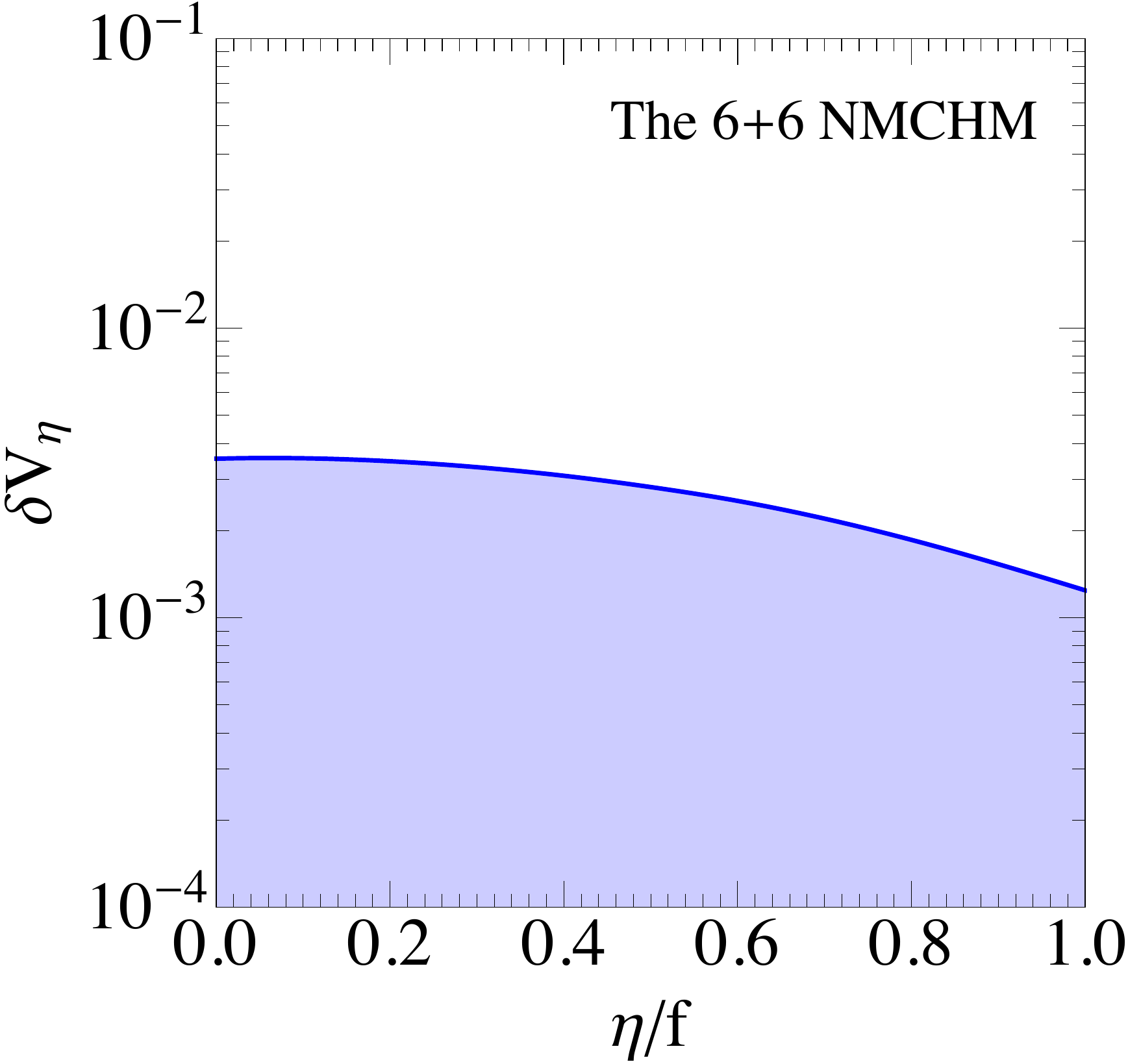}}\qquad
\subfigure{
\label{fig:F} 
\includegraphics[scale=0.385]{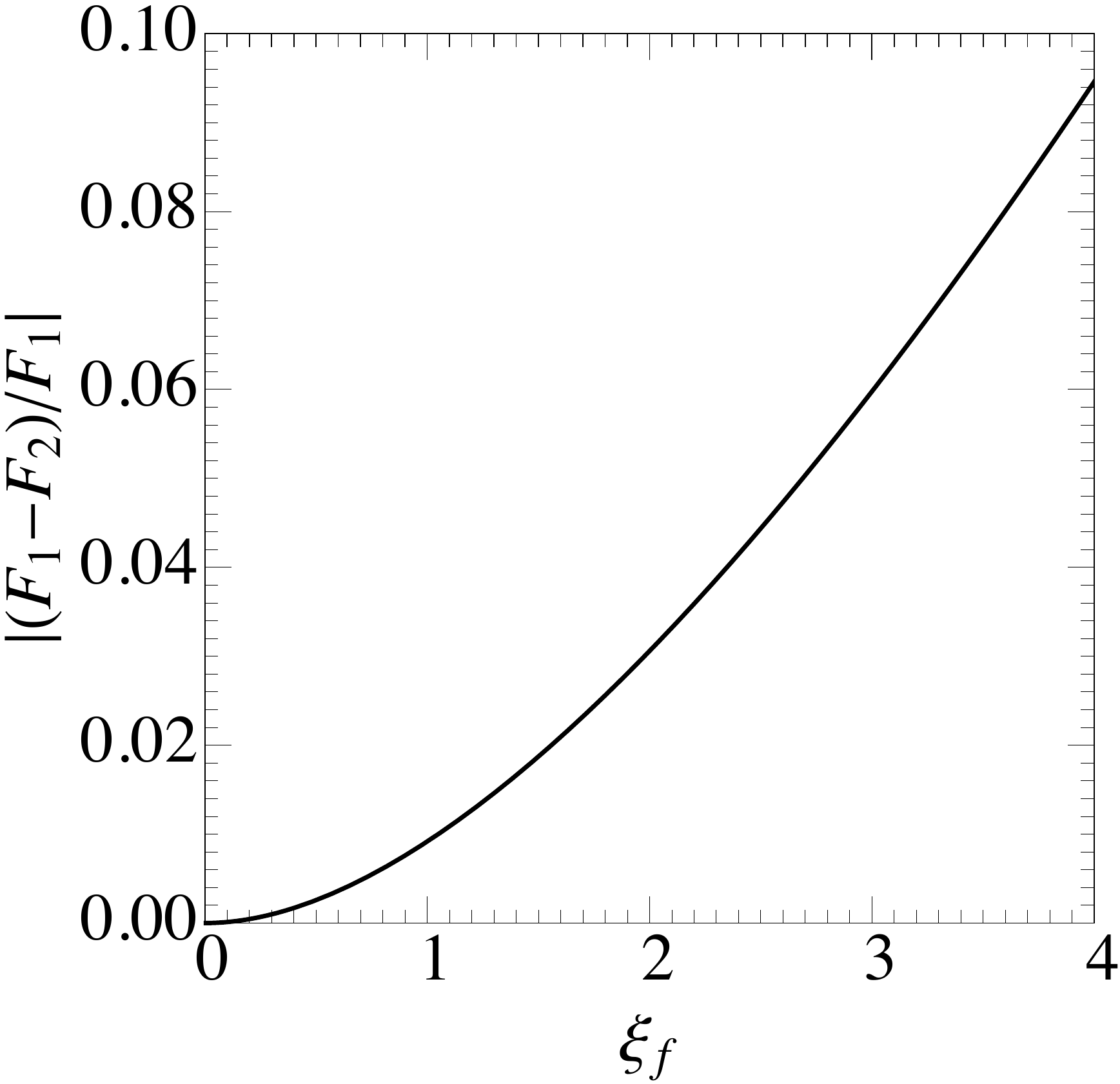}}
\caption{Left: the relative difference between potentials calculated by the complete Coleman-Weinberg integral and the polynomial expansion. Right: the relative difference of $F_1$ and $F_2$.}
\label{fig:polynomial}
\end{figure}

First we give a quantitative illustration about the difference between the two approaches (full calculation or polynomial expansion) using the SFOEWPT parameter points derived in Section~\ref{sec:6+6SFOEWPT}. The IR-driven potentials from those parameter points give $\abs{\eta}\gg f$, thus we need the UV contributions to enhance $\lambda_f$ to match the condition of SFOEWPT. Denoting the IR fully-calculated and polynomial potentials respectively as $V_{\rm full}^{\rm IR}$ and $V_{\rm poly}^{\rm IR}$, the relative difference of the full calculation and the polynomial expansion along the $\eta$ direction is defined as
\be
\delta V_\eta=\abs{\frac{V_{\rm full}^{\rm IR}(0,\eta)-V_{\rm poly}^{\rm IR}(0,\eta)}{V_{\rm full}^{\rm IR}(0,\eta)}},
\ee
which is a function of $\eta/f$. In left panel of Fig.~\ref{fig:polynomial} we show the envelope of $\delta V_\eta$ for all chosen parameter points. It can be read that the relative error of the polynomial approximation is within 0.4\%, even for $\eta/f\to1$. Thus the polynomial form of the potential is trustable. This is because of another additional factor in the logarithm expansion: $\Pi_1^{t}/\Pi_0^{t}$ (see \Eq{logarithm} for the details). Since $\Pi_1^{t}$ scales as $Q^{-6}$, $\Pi_1^{t}/\Pi_0^{t}$ is extremely small and the polynomial expansion works.

There is a semiquantitative way to understand the behavior of $V_{\rm full}^{\rm IR}(0,\eta)$. \Eq{logarithm} is an integral of type
\be\label{semi1}
V_{\rm full}^{\rm IR}(0,\eta)\sim F_1(\xi_f)=-2N_c\int\frac{Q^2dQ^2}{16\pi^2}\ln\left[1+\frac{\xi_f M^6}{\left(M^2+Q^2\right)^3}\right],
\ee
where $M\sim M_{\5,\1,\1'}$ and
\be
\xi_f\sim \frac{\abs{y_{L,R}^{\5,\1,\1'}}^2f^2}{M_{\5,\1,\1'}^2}\frac{\eta^2}{f^2}.
\ee
The polynomial expansion of $V_f$, on the other hand, is like
\be\label{semi2}
V_{\rm poly}^{\rm IR}(0,\eta)\sim F_2(\xi_f)=-2N_c\int\frac{Q^2dQ^2}{16\pi^2} \left[\frac{\xi_f M^6}{\left(M^2+Q^2\right)^3}-\frac12\left(\frac{\xi_f M^6}{\left(M^2+Q^2\right)^3}\right)^2\right].
\ee
The integrals in Eqs.~(\ref{semi1}) and (\ref{semi2}) can be analytically evaluated, and the relative difference $|(F_1-F_2)/F_1|$ depends only on $\xi_f$. The right panel of Fig.~\ref{fig:polynomial} shows this difference. One can see that even for $\xi_f=4$, the relative error is smaller than 9.5\%. Normally $\xi_f$ is at most $\mO(1)$, because we expect $|y_{L,R}^{\5,\1,\1'}|f\lesssim M_{\5,\1,\1'}$. For example, in the $\6+\6$ model $\xi_f<0.6$ for all points with SFOEWPT. Therefore, the relative difference of full expression and polynomial approximation is usually within 1\%.

\bibliographystyle{JHEP-2-2.bst}
\bibliography{references}

\providecommand{\href}[2]{#2}\begingroup\raggedright\begin{thebibliography}{10}

\bibitem{Kaplan:1991dc}
D.~B. Kaplan, ``{Flavor at SSC energies: A New mechanism for dynamically
  generated fermion
  masses},''\href{http://dx.doi.org/10.1016/S0550-3213(05)80021-5}{\emph{Nucl.
  Phys.} {\bf B365} (1991) 259--278}.

\bibitem{Contino:2003ve}
R.~Contino, Y.~Nomura and A.~Pomarol, ``{Higgs as a holographic pseudoGoldstone
  boson},''\href{http://dx.doi.org/10.1016/j.nuclphysb.2003.08.027}{\emph{Nucl.
  Phys.} {\bf B671} (2003) 148--174},
  [\href{https://arxiv.org/abs/hep-ph/0306259}{{\tt hep-ph/0306259}}].

\bibitem{Agashe:2004rs}
K.~Agashe, R.~Contino and A.~Pomarol, ``{The Minimal composite Higgs
  model},''\href{http://dx.doi.org/10.1016/j.nuclphysb.2005.04.035}{\emph{Nucl.
  Phys.} {\bf B719} (2005) 165--187},
  [\href{https://arxiv.org/abs/hep-ph/0412089}{{\tt hep-ph/0412089}}].

\bibitem{Contino:2006qr}
R.~Contino, L.~Da~Rold and A.~Pomarol, ``{Light custodians in natural composite
  Higgs
  models},''\href{http://dx.doi.org/10.1103/PhysRevD.75.055014}{\emph{Phys.
  Rev.} {\bf D75} (2007) 055014},
  [\href{https://arxiv.org/abs/hep-ph/0612048}{{\tt hep-ph/0612048}}].

\bibitem{Gripaios:2009pe}
B.~Gripaios, A.~Pomarol, F.~Riva and J.~Serra, ``{Beyond the Minimal Composite
  Higgs
  Model},''\href{http://dx.doi.org/10.1088/1126-6708/2009/04/070}{\emph{JHEP}
  {\bf 04} (2009) 070}, [\href{https://arxiv.org/abs/0902.1483}{{\tt
  0902.1483}}].

\bibitem{Cacciapaglia:2014uja}
G.~Cacciapaglia and F.~Sannino, ``{Fundamental Composite (Goldstone) Higgs
  Dynamics},''\href{http://dx.doi.org/10.1007/JHEP04(2014)111}{\emph{JHEP} {\bf
  04} (2014) 111}, [\href{https://arxiv.org/abs/1402.0233}{{\tt 1402.0233}}].

\bibitem{Barnard:2013zea}
J.~Barnard, T.~Gherghetta and T.~S. Ray, ``{UV descriptions of composite Higgs
  models without elementary
  scalars},''\href{http://dx.doi.org/10.1007/JHEP02(2014)002}{\emph{JHEP} {\bf
  02} (2014) 002}, [\href{https://arxiv.org/abs/1311.6562}{{\tt 1311.6562}}].

\bibitem{Ferretti:2013kya}
G.~Ferretti and D.~Karateev, ``{Fermionic UV completions of Composite Higgs
  models},''\href{http://dx.doi.org/10.1007/JHEP03(2014)077}{\emph{JHEP} {\bf
  03} (2014) 077}, [\href{https://arxiv.org/abs/1312.5330}{{\tt 1312.5330}}].

\bibitem{Hietanen:2014xca}
A.~Hietanen, R.~Lewis, C.~Pica and F.~Sannino, ``{Fundamental Composite Higgs
  Dynamics on the Lattice: SU(2) with Two
  Flavors},''\href{http://dx.doi.org/10.1007/JHEP07(2014)116}{\emph{JHEP} {\bf
  07} (2014) 116}, [\href{https://arxiv.org/abs/1404.2794}{{\tt 1404.2794}}].

\bibitem{Frigerio:2012uc}
M.~Frigerio, A.~Pomarol, F.~Riva and A.~Urbano, ``{Composite Scalar Dark
  Matter},''\href{http://dx.doi.org/10.1007/JHEP07(2012)015}{\emph{JHEP} {\bf
  07} (2012) 015}, [\href{https://arxiv.org/abs/1204.2808}{{\tt 1204.2808}}].

\bibitem{Marzocca:2014msa}
D.~Marzocca and A.~Urbano, ``{Composite Dark Matter and LHC
  Interplay},''\href{http://dx.doi.org/10.1007/JHEP07(2014)107}{\emph{JHEP}
  {\bf 07} (2014) 107}, [\href{https://arxiv.org/abs/1404.7419}{{\tt
  1404.7419}}].

\bibitem{Ma:2017vzm}
Y.~Wu, T.~Ma, B.~Zhang and G.~Cacciapaglia, ``{Composite Dark Matter and
  Higgs},''\href{http://dx.doi.org/10.1007/JHEP11(2017)058}{\emph{JHEP} {\bf
  11} (2017) 058}, [\href{https://arxiv.org/abs/1703.06903}{{\tt 1703.06903}}].

\bibitem{Cacciapaglia:2018avr}
G.~Cacciapaglia, S.~Vatani, T.~Ma and Y.~Wu, ``{Towards a fundamental safe
  theory of composite Higgs and Dark Matter},''
  \href{https://arxiv.org/abs/1812.04005}{{\tt 1812.04005}}.

\bibitem{Cai:2018tet}
C.~Cai, G.~Cacciapaglia and H.-H. Zhang, ``{Vacuum alignment in a composite
  2HDM},''\href{http://dx.doi.org/10.1007/JHEP01(2019)130}{\emph{JHEP} {\bf 01}
  (2019) 130}, [\href{https://arxiv.org/abs/1805.07619}{{\tt 1805.07619}}].

\bibitem{Bruggisser:2018mus}
S.~Bruggisser, B.~Von~Harling, O.~Matsedonskyi and G.~Servant, ``{Baryon
  Asymmetry from a Composite Higgs
  Boson},''\href{http://dx.doi.org/10.1103/PhysRevLett.121.131801}{\emph{Phys.
  Rev. Lett.} {\bf 121} (2018) 131801},
  [\href{https://arxiv.org/abs/1803.08546}{{\tt 1803.08546}}].

\bibitem{Bruggisser:2018mrt}
S.~Bruggisser, B.~Von~Harling, O.~Matsedonskyi and G.~Servant, ``{Electroweak
  Phase Transition and Baryogenesis in Composite Higgs
  Models},''\href{http://dx.doi.org/10.1007/JHEP12(2018)099}{\emph{JHEP} {\bf
  12} (2018) 099}, [\href{https://arxiv.org/abs/1804.07314}{{\tt 1804.07314}}].

\bibitem{Espinosa:2011eu}
J.~R. Espinosa, B.~Gripaios, T.~Konstandin and F.~Riva, ``{Electroweak
  Baryogenesis in Non-minimal Composite Higgs
  Models},''\href{http://dx.doi.org/10.1088/1475-7516/2012/01/012}{\emph{JCAP}
  {\bf 1201} (2012) 012}, [\href{https://arxiv.org/abs/1110.2876}{{\tt
  1110.2876}}].

\bibitem{Chala:2016ykx}
M.~Chala, G.~Nardini and I.~Sobolev, ``{Unified explanation for dark matter and
  electroweak baryogenesis with direct detection and gravitational wave
  signatures},''\href{http://dx.doi.org/10.1103/PhysRevD.94.055006}{\emph{Phys.
  Rev.} {\bf D94} (2016) 055006}, [\href{https://arxiv.org/abs/1605.08663}{{\tt
  1605.08663}}].

\bibitem{Chala:2018opy}
M.~Chala, M.~Ramos and M.~Spannowsky, ``{Gravitational wave and collider probes
  of a triplet Higgs sector with a low
  cutoff},''\href{http://dx.doi.org/10.1140/epjc/s10052-019-6655-1}{\emph{Eur.
  Phys. J.} {\bf C79} (2019) 156},
  [\href{https://arxiv.org/abs/1812.01901}{{\tt 1812.01901}}].

\bibitem{Audley:2017drz}
{\scshape LISA} collaboration, H.~Audley et~al., ``{Laser Interferometer Space
  Antenna},'' \href{https://arxiv.org/abs/1702.00786}{{\tt 1702.00786}}.

\bibitem{Luo:2015ght}
{\scshape TianQin} collaboration, J.~Luo et~al., ``{TianQin: a space-borne
  gravitational wave
  detector},''\href{http://dx.doi.org/10.1088/0264-9381/33/3/035010}{\emph{Class.
  Quant. Grav.} {\bf 33} (2016) 035010},
  [\href{https://arxiv.org/abs/1512.02076}{{\tt 1512.02076}}].

\bibitem{Hu:2017mde}
W.-R. Hu and Y.-L. Wu, ``{The Taiji Program in Space for gravitational wave
  physics and the nature of
  gravity},''\href{http://dx.doi.org/10.1093/nsr/nwx116}{\emph{Natl. Sci. Rev.}
  {\bf 4} (2017) 685--686}.

\bibitem{Crowder:2005nr}
J.~Crowder and N.~J. Cornish, ``{Beyond LISA: Exploring future gravitational
  wave
  missions},''\href{http://dx.doi.org/10.1103/PhysRevD.72.083005}{\emph{Phys.
  Rev.} {\bf D72} (2005) 083005},
  [\href{https://arxiv.org/abs/gr-qc/0506015}{{\tt gr-qc/0506015}}].

\bibitem{Kawamura:2011zz}
S.~Kawamura et~al., ``{The Japanese space gravitational wave antenna:
  DECIGO},''\href{http://dx.doi.org/10.1088/0264-9381/28/9/094011}{\emph{Class.
  Quant. Grav.} {\bf 28} (2011) 094011}.

\bibitem{Kawamura:2006up}
S.~Kawamura et~al., ``{The Japanese space gravitational wave antenna
  DECIGO},''\href{http://dx.doi.org/10.1088/0264-9381/23/8/S17}{\emph{Class.
  Quant. Grav.} {\bf 23} (2006) S125--S132}.

\bibitem{Mazumdar:2018dfl}
A.~Mazumdar and G.~White, ``{Review of cosmic phase transitions: their
  significance and experimental
  signatures},''\href{http://dx.doi.org/10.1088/1361-6633/ab1f55}{\emph{Rept.
  Prog. Phys.} {\bf 82} (2019) 076901},
  [\href{https://arxiv.org/abs/1811.01948}{{\tt 1811.01948}}].

\bibitem{Cline:2012hg}
J.~M. Cline and K.~Kainulainen, ``{Electroweak baryogenesis and dark matter
  from a singlet
  Higgs},''\href{http://dx.doi.org/10.1088/1475-7516/2013/01/012}{\emph{JCAP}
  {\bf 1301} (2013) 012}, [\href{https://arxiv.org/abs/1210.4196}{{\tt
  1210.4196}}].

\bibitem{Alanne:2014bra}
T.~Alanne, K.~Tuominen and V.~Vaskonen, ``{Strong phase transition, dark matter
  and vacuum stability from simple hidden
  sectors},''\href{http://dx.doi.org/10.1016/j.nuclphysb.2014.11.001}{\emph{Nucl.
  Phys.} {\bf B889} (2014) 692--711},
  [\href{https://arxiv.org/abs/1407.0688}{{\tt 1407.0688}}].

\bibitem{Huang:2015bta}
F.~P. Huang and C.~S. Li, ``{Electroweak baryogenesis in the framework of the
  effective field
  theory},''\href{http://dx.doi.org/10.1103/PhysRevD.92.075014}{\emph{Phys.
  Rev.} {\bf D92} (2015) 075014}, [\href{https://arxiv.org/abs/1507.08168}{{\tt
  1507.08168}}].

\bibitem{Vaskonen:2016yiu}
V.~Vaskonen, ``{Electroweak baryogenesis and gravitational waves from a real
  scalar
  singlet},''\href{http://dx.doi.org/10.1103/PhysRevD.95.123515}{\emph{Phys.
  Rev.} {\bf D95} (2017) 123515}, [\href{https://arxiv.org/abs/1611.02073}{{\tt
  1611.02073}}].

\bibitem{Huang:2017kzu}
F.~P. Huang and C.~S. Li, ``{Probing the baryogenesis and dark matter relaxed
  in phase transition by gravitational waves and
  colliders},''\href{http://dx.doi.org/10.1103/PhysRevD.96.095028}{\emph{Phys.
  Rev.} {\bf D96} (2017) 095028}, [\href{https://arxiv.org/abs/1709.09691}{{\tt
  1709.09691}}].

\bibitem{Huang:2018aja}
F.~P. Huang, Z.~Qian and M.~Zhang, ``{Exploring dynamical CP violation induced
  baryogenesis by gravitational waves and
  colliders},''\href{http://dx.doi.org/10.1103/PhysRevD.98.015014}{\emph{Phys.
  Rev.} {\bf D98} (2018) 015014}, [\href{https://arxiv.org/abs/1804.06813}{{\tt
  1804.06813}}].

\bibitem{Chiang:2018gsn}
C.-W. Chiang, Y.-T. Li and E.~Senaha, ``{Revisiting electroweak phase
  transition in the standard model with a real singlet
  scalar},''\href{http://dx.doi.org/10.1016/j.physletb.2018.12.017}{\emph{Phys.
  Lett.} {\bf B789} (2019) 154--159},
  [\href{https://arxiv.org/abs/1808.01098}{{\tt 1808.01098}}].

\bibitem{Bian:2018bxr}
L.~Bian and X.~Liu, ``{Two-step strongly first-order electroweak phase
  transition modified FIMP dark matter, gravitational wave signals, and the
  neutrino
  mass},''\href{http://dx.doi.org/10.1103/PhysRevD.99.055003}{\emph{Phys. Rev.}
  {\bf D99} (2019) 055003}, [\href{https://arxiv.org/abs/1811.03279}{{\tt
  1811.03279}}].

\bibitem{Bian:2018mkl}
L.~Bian and Y.-L. Tang, ``{Thermally modified sterile neutrino portal dark
  matter and gravitational waves from phase transition: The Freeze-in
  case},''\href{http://dx.doi.org/10.1007/JHEP12(2018)006}{\emph{JHEP} {\bf 12}
  (2018) 006}, [\href{https://arxiv.org/abs/1810.03172}{{\tt 1810.03172}}].

\bibitem{Cheng:2018axr}
W.~Cheng and L.~Bian, ``{Higgs inflation and cosmological electroweak phase
  transition with N scalars in the post-Higgs
  era},''\href{http://dx.doi.org/10.1103/PhysRevD.99.035038}{\emph{Phys. Rev.}
  {\bf D99} (2019) 035038}, [\href{https://arxiv.org/abs/1805.00199}{{\tt
  1805.00199}}].

\bibitem{Kurup:2017dzf}
G.~Kurup and M.~Perelstein, ``{Dynamics of Electroweak Phase Transition In
  Singlet-Scalar Extension of the Standard
  Model},''\href{http://dx.doi.org/10.1103/PhysRevD.96.015036}{\emph{Phys.
  Rev.} {\bf D96} (2017) 015036}, [\href{https://arxiv.org/abs/1704.03381}{{\tt
  1704.03381}}].

\bibitem{Contino:2010rs}
R.~Contino, ``{The Higgs as a Composite Nambu-Goldstone Boson},'' in
  \emph{{Physics of the large and the small, TASI 09, proceedings of the
  Theoretical Advanced Study Institute in Elementary Particle Physics, Boulder,
  Colorado, USA, 1-26 June 2009}}, pp.~235--306, 2011.
\newblock \href{https://arxiv.org/abs/1005.4269}{{\tt 1005.4269}}.
\newblock \href{http://dx.doi.org/10.1142/9789814327183_0005}{DOI}.

\bibitem{Shifman:1978bx}
M.~A. Shifman, A.~I. Vainshtein and V.~I. Zakharov, ``{QCD and Resonance
  Physics. Theoretical
  Foundations},''\href{http://dx.doi.org/10.1016/0550-3213(79)90022-1}{\emph{Nucl.
  Phys.} {\bf B147} (1979) 385--447}.

\bibitem{Shifman:1978by}
M.~A. Shifman, A.~I. Vainshtein and V.~I. Zakharov, ``{QCD and Resonance
  Physics:
  Applications},''\href{http://dx.doi.org/10.1016/0550-3213(79)90023-3}{\emph{Nucl.
  Phys.} {\bf B147} (1979) 448--518}.

\bibitem{Knecht:1997ts}
M.~Knecht and E.~de~Rafael, ``{Patterns of spontaneous chiral symmetry breaking
  in the large N(c) limit of QCD - like
  theories},''\href{http://dx.doi.org/10.1016/S0370-2693(98)00223-8}{\emph{Phys.
  Lett.} {\bf B424} (1998) 335--342},
  [\href{https://arxiv.org/abs/hep-ph/9712457}{{\tt hep-ph/9712457}}].

\bibitem{Marzocca:2012zn}
D.~Marzocca, M.~Serone and J.~Shu, ``{General Composite Higgs
  Models},''\href{http://dx.doi.org/10.1007/JHEP08(2012)013}{\emph{JHEP} {\bf
  08} (2012) 013}, [\href{https://arxiv.org/abs/1205.0770}{{\tt 1205.0770}}].

\bibitem{Pomarol:2012qf}
A.~Pomarol and F.~Riva, ``{The Composite Higgs and Light Resonance
  Connection},''\href{http://dx.doi.org/10.1007/JHEP08(2012)135}{\emph{JHEP}
  {\bf 08} (2012) 135}, [\href{https://arxiv.org/abs/1205.6434}{{\tt
  1205.6434}}].

\bibitem{Redi:2012ha}
M.~Redi and A.~Tesi, ``{Implications of a Light Higgs in Composite
  Models},''\href{http://dx.doi.org/10.1007/JHEP10(2012)166}{\emph{JHEP} {\bf
  10} (2012) 166}, [\href{https://arxiv.org/abs/1205.0232}{{\tt 1205.0232}}].

\bibitem{Banerjee:2017qod}
A.~Banerjee, G.~Bhattacharyya and T.~S. Ray, ``{Improving Fine-tuning in
  Composite Higgs
  Models},''\href{http://dx.doi.org/10.1103/PhysRevD.96.035040}{\emph{Phys.
  Rev.} {\bf D96} (2017) 035040}, [\href{https://arxiv.org/abs/1703.08011}{{\tt
  1703.08011}}].

\bibitem{Panico:2011pw}
G.~Panico and A.~Wulzer, ``{The Discrete Composite Higgs
  Model},''\href{http://dx.doi.org/10.1007/JHEP09(2011)135}{\emph{JHEP} {\bf
  09} (2011) 135}, [\href{https://arxiv.org/abs/1106.2719}{{\tt 1106.2719}}].

\bibitem{Coleman:1969sm}
S.~R. Coleman, J.~Wess and B.~Zumino, ``{Structure of phenomenological
  Lagrangians.
  1.},''\href{http://dx.doi.org/10.1103/PhysRev.177.2239}{\emph{Phys. Rev.}
  {\bf 177} (1969) 2239--2247}.

\bibitem{Callan:1969sn}
C.~G. Callan, Jr., S.~R. Coleman, J.~Wess and B.~Zumino, ``{Structure of
  phenomenological Lagrangians.
  2.},''\href{http://dx.doi.org/10.1103/PhysRev.177.2247}{\emph{Phys. Rev.}
  {\bf 177} (1969) 2247--2250}.

\bibitem{Panico:2015jxa}
G.~Panico and A.~Wulzer, ``{The Composite Nambu-Goldstone
  Higgs},''\href{http://dx.doi.org/10.1007/978-3-319-22617-0}{\emph{Lect. Notes
  Phys.} {\bf 913} (2016) pp.1--316},
  [\href{https://arxiv.org/abs/1506.01961}{{\tt 1506.01961}}].

\bibitem{Mrazek:2011iu}
J.~Mrazek, A.~Pomarol, R.~Rattazzi, M.~Redi, J.~Serra and A.~Wulzer, ``{The
  Other Natural Two Higgs Doublet
  Model},''\href{http://dx.doi.org/10.1016/j.nuclphysb.2011.07.008}{\emph{Nucl.
  Phys.} {\bf B853} (2011) 1--48}, [\href{https://arxiv.org/abs/1105.5403}{{\tt
  1105.5403}}].

\bibitem{Banerjee:2017wmg}
A.~Banerjee, G.~Bhattacharyya, N.~Kumar and T.~S. Ray, ``{Constraining
  Composite Higgs Models using LHC
  data},''\href{http://dx.doi.org/10.1007/JHEP03(2018)062}{\emph{JHEP} {\bf 03}
  (2018) 062}, [\href{https://arxiv.org/abs/1712.07494}{{\tt 1712.07494}}].

\bibitem{Espinosa:2011ax}
J.~R. Espinosa, T.~Konstandin and F.~Riva, ``{Strong Electroweak Phase
  Transitions in the Standard Model with a
  Singlet},''\href{http://dx.doi.org/10.1016/j.nuclphysb.2011.09.010}{\emph{Nucl.
  Phys.} {\bf B854} (2012) 592--630},
  [\href{https://arxiv.org/abs/1107.5441}{{\tt 1107.5441}}].

\bibitem{Dolan:1973qd}
L.~Dolan and R.~Jackiw, ``{Symmetry Behavior at Finite
  Temperature},''\href{http://dx.doi.org/10.1103/PhysRevD.9.3320}{\emph{Phys.
  Rev.} {\bf D9} (1974) 3320--3341}.

\bibitem{Braaten:1989kk}
E.~Braaten and R.~D. Pisarski, ``{Resummation and Gauge Invariance of the Gluon
  Damping Rate in Hot
  QCD},''\href{http://dx.doi.org/10.1103/PhysRevLett.64.1338}{\emph{Phys. Rev.
  Lett.} {\bf 64} (1990) 1338}.

\bibitem{Patel:2011th}
H.~H. Patel and M.~J. Ramsey-Musolf, ``{Baryon Washout, Electroweak Phase
  Transition, and Perturbation
  Theory},''\href{http://dx.doi.org/10.1007/JHEP07(2011)029}{\emph{JHEP} {\bf
  07} (2011) 029}, [\href{https://arxiv.org/abs/1101.4665}{{\tt 1101.4665}}].

\bibitem{Zhang:1992fs}
X.-m. Zhang, ``{Operators analysis for Higgs potential and cosmological bound
  on Higgs
  mass},''\href{http://dx.doi.org/10.1103/PhysRevD.47.3065}{\emph{Phys. Rev.}
  {\bf D47} (1993) 3065--3067},
  [\href{https://arxiv.org/abs/hep-ph/9301277}{{\tt hep-ph/9301277}}].

\bibitem{Grojean:2004xa}
C.~Grojean, G.~Servant and J.~D. Wells, ``{First-order electroweak phase
  transition in the standard model with a low
  cutoff},''\href{http://dx.doi.org/10.1103/PhysRevD.71.036001}{\emph{Phys.
  Rev.} {\bf D71} (2005) 036001},
  [\href{https://arxiv.org/abs/hep-ph/0407019}{{\tt hep-ph/0407019}}].

\bibitem{Gan:2017mcv}
X.~Gan, A.~J. Long and L.-T. Wang, ``{Electroweak sphaleron with dimension-six
  operators},''\href{http://dx.doi.org/10.1103/PhysRevD.96.115018}{\emph{Phys.
  Rev.} {\bf D96} (2017) 115018}, [\href{https://arxiv.org/abs/1708.03061}{{\tt
  1708.03061}}].

\bibitem{Huang:2015izx}
F.~P. Huang, P.-H. Gu, P.-F. Yin, Z.-H. Yu and X.~Zhang, ``{Testing the
  electroweak phase transition and electroweak baryogenesis at the LHC and a
  circular electron-positron
  collider},''\href{http://dx.doi.org/10.1103/PhysRevD.93.103515}{\emph{Phys.
  Rev.} {\bf D93} (2016) 103515}, [\href{https://arxiv.org/abs/1511.03969}{{\tt
  1511.03969}}].

\bibitem{Huang:2016odd}
F.~P. Huang, Y.~Wan, D.-G. Wang, Y.-F. Cai and X.~Zhang, ``{Hearing the echoes
  of electroweak baryogenesis with gravitational wave
  detectors},''\href{http://dx.doi.org/10.1103/PhysRevD.94.041702}{\emph{Phys.
  Rev.} {\bf D94} (2016) 041702}, [\href{https://arxiv.org/abs/1601.01640}{{\tt
  1601.01640}}].

\bibitem{Cao:2017oez}
Q.-H. Cao, F.~P. Huang, K.-P. Xie and X.~Zhang, ``{Testing the electroweak
  phase transition in scalar extension models at lepton
  colliders},''\href{http://dx.doi.org/10.1088/1674-1137/42/2/023103}{\emph{Chin.
  Phys.} {\bf C42} (2018) 023103},
  [\href{https://arxiv.org/abs/1708.04737}{{\tt 1708.04737}}].

\bibitem{Linde:1981zj}
A.~D. Linde, ``{Decay of the False Vacuum at Finite
  Temperature},''\href{http://dx.doi.org/10.1016/0550-3213(83)90293-6,
  10.1016/0550-3213(83)90072-X}{\emph{Nucl. Phys.} {\bf B216} (1983) 421}.

\bibitem{Moore:1998swa}
G.~D. Moore, ``{Measuring the broken phase sphaleron rate
  nonperturbatively},''\href{http://dx.doi.org/10.1103/PhysRevD.59.014503}{\emph{Phys.
  Rev.} {\bf D59} (1999) 014503},
  [\href{https://arxiv.org/abs/hep-ph/9805264}{{\tt hep-ph/9805264}}].

\bibitem{Quiros:1999jp}
M.~Quiros, ``{Finite temperature field theory and phase transitions},'' in
  \emph{{Proceedings, Summer School in High-energy physics and cosmology:
  Trieste, Italy, June 29-July 17, 1998}}, pp.~187--259, 1999.
\newblock \href{https://arxiv.org/abs/hep-ph/9901312}{{\tt hep-ph/9901312}}.

\bibitem{Zhou:2019uzq}
R.~Zhou, L.~Bian and H.-K. Guo, ``{Probing the Electroweak Sphaleron with
  Gravitational Waves},'' \href{https://arxiv.org/abs/1910.00234}{{\tt
  1910.00234}}.

\bibitem{Denner:2011mq}
A.~Denner, S.~Heinemeyer, I.~Puljak, D.~Rebuzzi and M.~Spira, ``{Standard Model
  Higgs-Boson Branching Ratios with
  Uncertainties},''\href{http://dx.doi.org/10.1140/epjc/s10052-011-1753-8}{\emph{Eur.
  Phys. J.} {\bf C71} (2011) 1753},
  [\href{https://arxiv.org/abs/1107.5909}{{\tt 1107.5909}}].

\bibitem{ATLAS-CONF-2018-031}
{\scshape ATLAS Collaboration} collaboration, ``{Combined measurements of Higgs
  boson production and decay using up to 80 fb$^{-1}$ of proton--proton
  collision data at $\sqrt{s}=$ 13 TeV collected with the ATLAS experiment},''
  Tech. Rep. ATLAS-CONF-2018-031, CERN, Geneva, Jul, 2018.

\bibitem{Sirunyan:2018koj}
{\scshape CMS} collaboration, A.~M. Sirunyan et~al., ``{Combined measurements
  of Higgs boson couplings in proton–proton collisions at $\sqrt{s}=13\,\text
  {Te}\text {V}
  $},''\href{http://dx.doi.org/10.1140/epjc/s10052-019-6909-y}{\emph{Eur. Phys.
  J.} {\bf C79} (2019) 421}, [\href{https://arxiv.org/abs/1809.10733}{{\tt
  1809.10733}}].

\bibitem{Georgi:1986df}
H.~Georgi, D.~B. Kaplan and L.~Randall, ``{Manifesting the Invisible Axion at
  Low-energies},''\href{http://dx.doi.org/10.1016/0370-2693(86)90688-X}{\emph{Phys.
  Lett.} {\bf 169B} (1986) 73--78}.

\bibitem{Jackiw:1974cv}
R.~Jackiw, ``{Functional evaluation of the effective
  potential},''\href{http://dx.doi.org/10.1103/PhysRevD.9.1686}{\emph{Phys.
  Rev.} {\bf D9} (1974) 1686}.

\bibitem{Weinberg:1967kj}
S.~Weinberg, ``{Precise relations between the spectra of vector and axial
  vector
  mesons},''\href{http://dx.doi.org/10.1103/PhysRevLett.18.507}{\emph{Phys.
  Rev. Lett.} {\bf 18} (1967) 507--509}.

\bibitem{Feroz:2008xx}
F.~Feroz, M.~P. Hobson and M.~Bridges, ``{MultiNest: an efficient and robust
  Bayesian inference tool for cosmology and particle
  physics},''\href{http://dx.doi.org/10.1111/j.1365-2966.2009.14548.x}{\emph{Mon.
  Not. Roy. Astron. Soc.} {\bf 398} (2009) 1601--1614},
  [\href{https://arxiv.org/abs/0809.3437}{{\tt 0809.3437}}].

\bibitem{ALEPH:2005ab}
{\scshape ALEPH, DELPHI, L3, OPAL, SLD, LEP Electroweak Working Group, SLD
  Electroweak Group, SLD Heavy Flavour Group} collaboration, S.~Schael et~al.,
  ``{Precision electroweak measurements on the $Z$
  resonance},''\href{http://dx.doi.org/10.1016/j.physrep.2005.12.006}{\emph{Phys.
  Rept.} {\bf 427} (2006) 257--454},
  [\href{https://arxiv.org/abs/hep-ex/0509008}{{\tt hep-ex/0509008}}].

\bibitem{Tanabashi:2018oca}
{\scshape Particle Data Group} collaboration, M.~Tanabashi et~al., ``{Review of
  Particle
  Physics},''\href{http://dx.doi.org/10.1103/PhysRevD.98.030001}{\emph{Phys.
  Rev.} {\bf D98} (2018) 030001}.

\bibitem{Wainwright:2011kj}
C.~L. Wainwright, ``{CosmoTransitions: Computing Cosmological Phase Transition
  Temperatures and Bubble Profiles with Multiple
  Fields},''\href{http://dx.doi.org/10.1016/j.cpc.2012.04.004}{\emph{Comput.
  Phys. Commun.} {\bf 183} (2012) 2006--2013},
  [\href{https://arxiv.org/abs/1109.4189}{{\tt 1109.4189}}].

\bibitem{Grojean:2006bp}
C.~Grojean and G.~Servant, ``{Gravitational Waves from Phase Transitions at the
  Electroweak Scale and
  Beyond},''\href{http://dx.doi.org/10.1103/PhysRevD.75.043507}{\emph{Phys.
  Rev.} {\bf D75} (2007) 043507},
  [\href{https://arxiv.org/abs/hep-ph/0607107}{{\tt hep-ph/0607107}}].

\bibitem{Caprini:2015zlo}
C.~Caprini et~al., ``{Science with the space-based interferometer eLISA. II:
  Gravitational waves from cosmological phase
  transitions},''\href{http://dx.doi.org/10.1088/1475-7516/2016/04/001}{\emph{JCAP}
  {\bf 1604} (2016) 001}, [\href{https://arxiv.org/abs/1512.06239}{{\tt
  1512.06239}}].

\bibitem{Steinhardt:1981ct}
P.~J. Steinhardt, ``{Relativistic Detonation Waves and Bubble Growth in False
  Vacuum Decay},''\href{http://dx.doi.org/10.1103/PhysRevD.25.2074}{\emph{Phys.
  Rev.} {\bf D25} (1982) 2074}.

\bibitem{Ellis:2018mja}
J.~Ellis, M.~Lewicki and J.~M. No, ``{On the Maximal Strength of a First-Order
  Electroweak Phase Transition and its Gravitational Wave Signal},''
  \href{https://arxiv.org/abs/1809.08242}{{\tt 1809.08242}}.

\bibitem{Franzosi:2016aoo}
D.~Buarque~Franzosi, G.~Cacciapaglia, H.~Cai, A.~Deandrea and M.~Frandsen,
  ``{Vector and Axial-vector resonances in composite models of the Higgs
  boson},''\href{http://dx.doi.org/10.1007/JHEP11(2016)076}{\emph{JHEP} {\bf
  11} (2016) 076}, [\href{https://arxiv.org/abs/1605.01363}{{\tt 1605.01363}}].

\bibitem{Niehoff:2016zso}
C.~Niehoff, P.~Stangl and D.~M. Straub, ``{Electroweak symmetry breaking and
  collider signatures in the next-to-minimal composite Higgs
  model},''\href{http://dx.doi.org/10.1007/JHEP04(2017)117}{\emph{JHEP} {\bf
  04} (2017) 117}, [\href{https://arxiv.org/abs/1611.09356}{{\tt 1611.09356}}].

\bibitem{Liu:2018hum}
D.~Liu, L.-T. Wang and K.-P. Xie, ``{Prospects of searching for composite
  resonances at the LHC and
  beyond},''\href{http://dx.doi.org/10.1007/JHEP01(2019)157}{\emph{JHEP} {\bf
  01} (2019) 157}, [\href{https://arxiv.org/abs/1810.08954}{{\tt 1810.08954}}].

\bibitem{Aad:2019fbh}
{\scshape ATLAS} collaboration, G.~Aad et~al., ``{Search for diboson resonances
  in hadronic final states in 139 fb$^{-1}$ of $pp$ collisions at $\sqrt{s} =
  13$ TeV with the ATLAS detector},''
  \href{https://arxiv.org/abs/1906.08589}{{\tt 1906.08589}}.

\bibitem{Sirunyan:2018yun}
{\scshape CMS} collaboration, A.~M. Sirunyan et~al., ``{Search for top quark
  partners with charge 5/3 in the same-sign dilepton and single-lepton final
  states in proton-proton collisions at $ \sqrt{s}=13 $
  TeV},''\href{http://dx.doi.org/10.1007/JHEP03(2019)082}{\emph{JHEP} {\bf 03}
  (2019) 082}, [\href{https://arxiv.org/abs/1810.03188}{{\tt 1810.03188}}].

\bibitem{Kim:2019rhy}
J.~H. Kim, K.~Kong, B.~Nachman and D.~Whiteson, ``{The motivation and status of
  two-body resonance decays after the LHC Run 2 and beyond},''
  \href{https://arxiv.org/abs/1907.06659}{{\tt 1907.06659}}.

\bibitem{Barducci:2015vyf}
D.~Barducci and C.~Delaunay, ``{Bounding wide composite vector resonances at
  the LHC},''\href{http://dx.doi.org/10.1007/JHEP02(2016)055}{\emph{JHEP} {\bf
  02} (2016) 055}, [\href{https://arxiv.org/abs/1511.01101}{{\tt 1511.01101}}].

\bibitem{Vignaroli:2014bpa}
N.~Vignaroli, ``{New W' signals at the
  LHC},''\href{http://dx.doi.org/10.1103/PhysRevD.89.095027}{\emph{Phys. Rev.}
  {\bf D89} (2014) 095027}, [\href{https://arxiv.org/abs/1404.5558}{{\tt
  1404.5558}}].

\bibitem{Cacciapaglia:2015eqa}
G.~Cacciapaglia, H.~Cai, A.~Deandrea, T.~Flacke, S.~J. Lee and A.~Parolini,
  ``{Composite scalars at the LHC: the Higgs, the Sextet and the
  Octet},''\href{http://dx.doi.org/10.1007/JHEP11(2015)201}{\emph{JHEP} {\bf
  11} (2015) 201}, [\href{https://arxiv.org/abs/1507.02283}{{\tt 1507.02283}}].

\bibitem{Belyaev:2016ftv}
A.~Belyaev, G.~Cacciapaglia, H.~Cai, G.~Ferretti, T.~Flacke, A.~Parolini
  et~al., ``{Di-boson signatures as Standard Candles for Partial
  Compositeness},''\href{http://dx.doi.org/10.1007/JHEP01(2017)094,
  10.1007/JHEP12(2017)088}{\emph{JHEP} {\bf 01} (2017) 094},
  [\href{https://arxiv.org/abs/1610.06591}{{\tt 1610.06591}}].

\bibitem{Serra:2015xfa}
J.~Serra, ``{Beyond the Minimal Top Partner
  Decay},''\href{http://dx.doi.org/10.1007/JHEP09(2015)176}{\emph{JHEP} {\bf
  09} (2015) 176}, [\href{https://arxiv.org/abs/1506.05110}{{\tt 1506.05110}}].

\bibitem{Bizot:2018tds}
N.~Bizot, G.~Cacciapaglia and T.~Flacke, ``{Common exotic decays of top
  partners},''\href{http://dx.doi.org/10.1007/JHEP06(2018)065}{\emph{JHEP} {\bf
  06} (2018) 065}, [\href{https://arxiv.org/abs/1803.00021}{{\tt 1803.00021}}].

\bibitem{Xie:2019gya}
K.-P. Xie, G.~Cacciapaglia and T.~Flacke, ``{Exotic decays of top partners with
  charge 5/3: bounds and opportunities},''
  \href{https://arxiv.org/abs/1907.05894}{{\tt 1907.05894}}.

\bibitem{Cacciapaglia:2019zmj}
G.~Cacciapaglia, T.~Flacke, M.~Park and M.~Zhang, ``{Exotic decays of top
  partners: mind the search gap},''
  \href{https://arxiv.org/abs/1908.07524}{{\tt 1908.07524}}.

\bibitem{No:2011fi}
J.~M. No, ``{Large Gravitational Wave Background Signals in Electroweak
  Baryogenesis
  Scenarios},''\href{http://dx.doi.org/10.1103/PhysRevD.84.124025}{\emph{Phys.
  Rev.} {\bf D84} (2011) 124025}, [\href{https://arxiv.org/abs/1103.2159}{{\tt
  1103.2159}}].

\bibitem{Alves:2018oct}
A.~Alves, T.~Ghosh, H.-K. Guo and K.~Sinha, ``{Resonant Di-Higgs Production at
  Gravitational Wave Benchmarks: A Collider Study using Machine
  Learning},''\href{http://dx.doi.org/10.1007/JHEP12(2018)070}{\emph{JHEP} {\bf
  12} (2018) 070}, [\href{https://arxiv.org/abs/1808.08974}{{\tt 1808.08974}}].

\bibitem{Alves:2018jsw}
A.~Alves, T.~Ghosh, H.-K. Guo, K.~Sinha and D.~Vagie, ``{Collider and
  Gravitational Wave Complementarity in Exploring the Singlet Extension of the
  Standard
  Model},''\href{http://dx.doi.org/10.1007/JHEP04(2019)052}{\emph{JHEP} {\bf
  04} (2019) 052}, [\href{https://arxiv.org/abs/1812.09333}{{\tt 1812.09333}}].

\bibitem{Agashe:2006at}
K.~Agashe, R.~Contino, L.~Da~Rold and A.~Pomarol, ``{A Custodial symmetry for
  $Zb \bar
  b$},''\href{http://dx.doi.org/10.1016/j.physletb.2006.08.005}{\emph{Phys.
  Lett.} {\bf B641} (2006) 62--66},
  [\href{https://arxiv.org/abs/hep-ph/0605341}{{\tt hep-ph/0605341}}].

\bibitem{Gori:2015nqa}
S.~Gori, J.~Gu and L.-T. Wang, ``{The $ Zb\overline{b} $ couplings at future
  e$^{+}$ e$^{-}$
  colliders},''\href{http://dx.doi.org/10.1007/JHEP04(2016)062}{\emph{JHEP}
  {\bf 04} (2016) 062}, [\href{https://arxiv.org/abs/1508.07010}{{\tt
  1508.07010}}].

\end{thebibliography}\endgroup

\end{document}